\documentclass[12pt]{iopart}

\usepackage{comment}
\usepackage{csquotes}
\usepackage{amsthm}
\usepackage{amsfonts}
\usepackage{multirow}
\usepackage[hidelinks]{hyperref}

\usepackage{graphicx}% Include figure files
\usepackage{tikz}
\usetikzlibrary{shapes.geometric}
\usepackage{dcolumn}% Align table columns on decimal point
\usepackage{bm}% bold math
\usepackage{xcolor}

\definecolor{edits}{HTML}{000000}
\newtheorem{theorem}{Theorem}

\usepackage{appendix}
\usepackage[normalem]{ulem}

\usepackage[normalem]{ulem}

\newcommand{\LL}{\mathcal{L}}

\usepackage{tikz}
\usepackage{adjustbox}
\usetikzlibrary{quantikz2}

\usepackage{mathtools}

\begin{document}

\title{{\color{edits} Gaussianity and} Simulability of Cliffords and Matchgates}% Force line breaks with \\

\author{Andrew M Projansky$^1$, Jason Necaise$^1$, James D Whitfield$^{1,2}$}

\address{$^1$ Department of Physics and Astronomy, Dartmouth College, Hanover, New Hampshire, USA 03755}
\address{$^2$ AWS Center for Quantum Computing, Pasadena, California, USA 91125}

\ead{andrew.m.projansky.gr@dartmouth.edu}

\date{\today}% It is always \today, today,


\date{\today}% It is always \today, today,

\begin{abstract}
Though Cliffords and matchgates are both examples of classically simulable circuits, they are considered simulable for different reasons. {\color{edits} The celebrated Gottesman-Knill explains the simulability Cliffords, and the efficient simulability of matchgates is understood via Pfaffians of antisymmetric matrices. We take the perspective that by studying Clifford-matchgate hybrid circuits, we expand the set of known simulable circuits and reach a better understanding of what unifies these two circuit families}. While the simulability of Clifford conjugated matchgate circuits for single qubit outputs has been briefly considered, the simulability of Clifford and matchgate hybrid circuits has not been generalized up to this point. In this paper {\color{edits} we extend that work}, studying simulability of marginals as well as Pauli expectation values of Clifford and matchgate hybrid circuits. We describe a hierarchy of Clifford circuits, and find that as we consider more general Cliffords, we lose some amount of simulability of bitstring outputs. We then show that the known simulability of Pauli expectation values of Clifford circuits acting on product states can be generalized to Clifford circuits acting after any matchgate circuit. We conclude with general discussion about the relationship between Cliffords and matchgates, {\color{edits} and show that both circuit families can be understood as being Gaussian. }
\end{abstract}

\maketitle

\tableofcontents
\newpage

\section{\label{sec:Intro}Introduction}
%The line break was forced \lowercase{via} \textbackslash\textbackslash}

It is well known that %just because we have 
access to a quantum computer does not mean we have access now to 
efficient solvability of all problems. For example, there are problems like solving for the ground states of arbitrary %molecular 
Hamiltonians which, in general, are not efficiently solvable on a quantum computer \cite{Whitfield_2013, aaronson2009bqp}. We also have problems which naively may appear to provide quantum advantage, but can really be simulated on a classical computer in polynomial time. Some examples of these kinds of problems are those relating to weakly entangling circuits, Clifford circuits and matchgate circuits. Note, here we use the term simulable to refer to classically efficient simulability. \\
\indent Clifford circuits, well studied in their connection to quantum error correction, define a class of simulable circuits due to the fact that the Cliffords are the normalizers of the Pauli group \cite{nielsen00, Aaronson_2004, gottesmanknill, JoszaCliffC, Koh_2017}. Matchgates, though first introduced by Valiant \cite{valiant, VALIANT2002457} in the context of the perfect matching problem on planar graphs, are {\color{edits}often} discussed with their connection to free fermion time dynamics. Here, ``free'' refers to free from interactions. Terhal, Divincenzo \cite{TerhalDivincenzo} and Knill \cite{knill2001fermionic} found that matchgates could be understood by mapping the generating Hamiltonians back to free fermions, using the Jordan-Wigner mapping \cite{1928ZPhy...47..631J}. \\
\indent Though both Clifford and matchgate circuits are simulable, their simulability is often considered separately. While general operator structure and group structure have been studied to describe both, the proofs of simulability for each are different \cite{jozsa2015jordanwigner, ermakov2024unifiedframeworkefficientlycomputable}. The tools used to classically simulate each system are also different. One relies on Clifford tableaux, while the other often is connected to the efficient computation of Pfaffians of antisymmetric matrices \cite{Aaronson_2004, TerhalDivincenzo}. \\
\indent In this paper we {\color{edits} study the ways in which we can combine Cliffords and matchgates together while retaining classical simulability. This work accomplishes three different tasks. We classify the complexity of conjugated matchgates and the difficulty of calculating probability marginals in these hybrid circuits. We are able to then extend the simulability of matchgates and Cliffords by considering the action of Cliffords following matchgates. Finally, we are able to discuss how we can understand stabilizers states and Gaussian states from the same perspective, thus bridging the gap between the two seemingly separate systems. In doing so we sharpen the divide between quantum and classical, leading to a better understanding the structure of simulable circuits while specifying what resources enable the jump from classical to quantum computing. As we search for applications of quantum computing to real world problems, this quantum/classical border allows us to narrow our search while introducing new classical approximations we can use to study quantum systems today. } \\
\indent While briefly considered {\color{edits} at} the end of work from Jozsa and Miyake \cite{JozsaMiyake} and in work from Van den Nest \cite{nest2010simulatingquantumcomputersprobabilistic}, this is the first in depth study of the simulability of various Clifford and matchgate hybrid circuits. 
{\color{edits} We present results in the main body of this work. Technical details and relevant background can be found in later sections.} \\

\section{Results}

\subsection{Definitions for Results}

{\color{edits} We review a series of abbreviated definitions needed to present the theorems proven in this work. We define all terms used more formally in later sections. \\ \\
\noindent \textit{Definition A. Cliffords}: a discrete set of gates generated by the Hadamard, CNOT, and Phase gates. We note that the SWAP and CZ gates are also Clifford gates. \\ \\
\noindent \textit{Definition B. Fermion-to-Qubit encoding}: an ordered set $\mathcal{L}_1^{(n)}$ of mutually anti-commuting Pauli strings representing Majorana fermionic operators with respect to spin systems:
\begin{equation}
    \mathcal{L}_1^{(n)} \equiv \{ \ c_i \ | \ c_i \in \mathcal{P}_n, \ \{c_i, c_j\} = 2\delta_{ij} \ \}.
\end{equation}
$\mathcal{L}_1^{(n)}$ represents the Majoranas of some encoding, and $\mathcal{L}_{1,2}^{(n)}$ represents those Majoranas \textit{and} quadratic products of those Majoranas. \\

\noindent \textit{Definition C. Matchgates}: a continuous family of parity preserving unitary gates $U_{\text{MG}} = e^{-iH}$ for $H = \sum_{i,j}^{2n}h_{ij}c_i c_j$, the dynamics of a fermion Hamiltonian written with the Jordan-Wigner fermion-to-qubit encoding.
\\ \\
\noindent \textit{Definition D. Inputs/Outputs}: the initial states of the circuits, and the desired quantity we want to calculate, respectively. For inputs we consider  product state inputs ($PI$) or computational basis state inputs in the $Z$ basis ($CI$). For outputs we consider the following. 
\begin{itemize}
    \item Bitstring outputs ($BO$), i.e, the probability of measuring some subset of qubits $k$ to be in some bitstring  $y^*$ over $k$-bits. 
    \item Pauli outputs ($PO$), i.e, the expectation value of some Pauli $p$. 
\end{itemize}
\vspace{0.35cm}
\noindent \textit{Definition E. Gaussianity}: a state $|\psi\rangle$ is Gaussian if and only if \cite{bravyi2004lagrangianrepresentationfermioniclinear}

\begin{equation*}
    \sum_{i=1}^{2n} c_k \otimes c_k |\psi\rangle |\psi\rangle = 0 {\color{edits}.}
\end{equation*}
}
\subsection{Conjugated Matchgates, and Fermionic Encodings} 

{\color{edits} We prove a series of theorems specifying the complexity of conjugated matchgate circuits, via reduction to matchgate problems. While proof techniques appear simple, the value of our work comes from how we understand the complexity introduced via these conjugations, and the relation to fermion-to-qubit transformations.  While not directly providing new simulable circuits, we can understand our results both from the perspective of complexity theory and from physical intuition about the ways in which we attempt to represent fermionic information in qubit systems. We are also able to introduce constructions of magic state resources for matchgate computations that utilize restricted Clifford resources. \\
\indent The starting point is the well known Jordan-Wigner encoding; Unitaries generated from free fermion Hamiltonians represented with spins via the Jordan-Wigner encoding are exactly matchgate circuits, for which we have known simulability results for the past ~25 years. At a higher level, the Jordan-Wigner encoding is special; it} is the encoding which encodes information {\color{edits} from Fock Space} in the exact same way in qubit space. For the Jordan-Wigner encoding, the occupancy information of a single mode in Fock space is mapped to a single qubit site being $1$ or $0$. We also have that all Fock basis states are mapped exactly to the same state in qubit space: as an example, the Fock state $|0010001110\rangle$ is mapped to qubit state $|0010001110\rangle$. \\
\indent For other encodings, this is not true. Take for example the Bravyi-Kitaev encoding or Ternary Tree encoding, where {\color{edits}the encoding's} information about the occupancy of one fermionic mode can be stored in a $O(\log(n))$ scaling number of qubits \cite{Setia_2018, Jiang_2020}.  \\
\indent We can understand {\color{edits} the} hierarchy of Clifford conjugated circuits as different ways of representing fermionic information in qubit space, and see that the simulability of Clifford conjugated circuits directly corresponds to how close {\color{edits} the} fermionic information represented in qubit space looks like its Fock space representation. \\
\indent For Jordan-Wigner and SWAP conjugated Jordan-Wigner, for which we retain the most simulable outputs, we are encoding Fock space information equivalently in qubit space to how it looks in Fock space.\\
\indent For $CZ$ conjugated matchgates, occupancy information is stored the same, but we find that the way Fock states get encoded into qubit states is slightly different. As an example, a Fock state like $|0010001110\rangle$ may be mapped to qubit state $-|0010001110\rangle$. This information being encoded into phases is enough that while we retain simulability on computational basis states, we are no longer able to simulate over product states. \\
\indent For Clifford permutation conjugated matchgates, we are dealing with encodings which specifically map Fock basis states to $Z$ basis states, though we now can have information of the occupancy of one fermionic mode stored across multiple qubits. This spreading of occupancy information correlates with the loss of calculating marginals efficiently. \\
\indent At the bottom level of {\color{edits} the} hierarchy, for arbitrary Clifford {\color{edits}conjugation} we have that {\color{edits} the} information from Fock space may be encoded highly non-locally in qubit space, with {\color{edits} the} basis states stored across possibly entangled superpositions of qubit states. {\color{edits} In summary,} as the representation of fermionic information in qubit space deviates further and further from the information in Fock space, we lose more and more simulability. \\
\indent {\color{edits} We now present a series of theorems which highlight the complexity of conjugated matchgates and their associated fermionic encodings \cite{ConjbCliff}. We note specifically that we are focused on what in the literature is called \textit{strong} simulation; the ability to calculate $k$-bit probability marginals. We point out that \textit{strong} simulability implies \textit{weak} simulability, or the ability to efficiently generate samples \cite{terhal2004adaptivequantumcomputationconstant}}.  \\

\begin{theorem}\label{JWPS} {\color{edits} SWAP Conjugated Matchgates.}
    \begin{enumerate}
        \item {\color{edits} SWAP conjugated matchgates are PIBO simulable (Product-input bitstring-output). }
        \item The only ancilla-free encodings such that the generators required to make arbitrary product state are within $\mathcal{L}_{1,2}^{(n)}$ are the Jordan-Wigner encoding {\color{edits} SWAP conjugated Jordan-Wigner encoding. There is also the freedom to conjugate by a single local Clifford on one specified qubit}.  
    \end{enumerate}
\end{theorem}

{\color{edits} The proof of statement 1 is trivial, and can be done with classical processing of the product state input and bitstring output. \\
\indent The proof of statement 2 is far less trivial. While product states are simple from the perspective of states in qubit space, we prove that outside of Jordan-Wigner product states require fermionic resources. Thinking of product states in the context of these conjugated circuits gives us ways of combining independently simulable resources in computationally complex ways.} \\

\begin{theorem}\label{thm:CZM}
    {\color{edits} CZ+SWAP conjugated matchgate circuits}
    \begin{enumerate}
        \item  {\color{edits} CZ+SWAP conjugated matchgate circuits are CIBO (Computational input, bitstring output) simulable.}
        \item If $Z_i\in \mathcal{L}_{2}^{(n)}$ for all $i$, then the ancilla-free encoding is connected to Jordan-Wigner by a circuit composed of only SWAP and CZ gates (we will denote the family of Cliffords generated by SWAPs and $CZ$ gates be defined as $C_{\text{CZ}}$).
        \item Simulability of bitstring outputs of $CZ$ conjugated matchgate circuits on product state inputs {\color{edits}implies} the ability to simulate bitstring outputs of arbitrary quantum circuits {\color{edits} and a collapse of the polynomial hierarchy to its third level. }
    \end{enumerate}
\end{theorem}
{\color{edits} Statement 1, again, is trivial with classical processing. Statements 2 and 3 are novel. In proving statement 2, we show that requiring $Z_i\in \mathcal{L}_{2}^{(n)}$ for all $i$ places heavy restrictions on {\color{edits} the} possible fermionic encodings, but still allows for interesting encodings where fermionic occupancy is still stored locally. While statement 3 follows from Theorem \hyperref[JWPS]{1}, we specify that simulability would collapse the polynomial hierarchy to its third level \cite{terhal2004adaptivequantumcomputationconstant} via magic state gadget construction, following \cite{MGMagic}. Not only are product states not Gaussian for these conjugated circuits, but they are resourceful states that would allow for universal quantum computation.} 

\begin{theorem}\label{thm:CPC} 
    For any Clifford permutation conjugated matchgate circuit, we have at least efficient CIbO (computational input, restricted bitstring output) simulation.
\end{theorem}

{\color{edits} Unlike the statements above, the mapping from conjugated to unconjugated matchgate problems for Clifford permutations is not trivial. Specifically, the only time a reduction to a matchgate problem exist is when we place a restriction that output bitstrings are over the entire system.}

\begin{theorem}
    For an arbitrary Clifford $C$, {\color{edits} reduction to a matchgate problem does not necessarily exist. }
\end{theorem}

{\color{edits} We present an example in which no reduction to a matchgate problem exist. We conjecture that efficient simulability would imply simulability of universal circuits, though a strict proof will hopefully be included in future work.}

\subsection{Pauli Expectation Values, Covariance Matrices, and Cliffords following Matchgates}

{\color{edits} We turn now to focus on the ability to compute Pauli expectation values efficiently for Clifford and matchgate hybrid circuits. This work is an extension of Jozsa and Miyake, \cite{JozsaMiyake} in which Pauli expectation values are of interest. We utilize the fact that Gaussian fermionic states/matchgate circuits can be represented via covariance matrices, matrices which keep track of expectation values over quadratic products of majoranas. We notice that Cliffords only change the algebra the covariance matrix describes, rather than the covariance matrix itself. This result can be seen as an extension of the simulability of Pauli expectation values for Cliffords on product states. \\
\indent This result also ties together Cliffords and matchgates in a novel way. Clifford and matchgate simulability are often treated as two completely separate ways. We show that covariance matrices can be used to compute Pauli expectation values over both circuit families; thus, a single object allows us to represent both circuit families. We can thus consider that the simulability of stabilizer states can be understood due to their Gaussian structure.}

\begin{theorem}\label{thm:CIPO}
    For arbitrary Cliffords $C$, for any circuit of the form $C^{\dagger}U_{\text{MG}}$, we have at least efficient CIPO simulation {\color{edits} via $2n \times 2n$ covariance matrices}.
\end{theorem}

{\color{edits} A similar but more powerful family of circuits is proven to be simulable below. }

\begin{theorem}\label{thm:PIPO}
    For arbitrary Cliffords $C$, for any circuit of the form $C^{\dagger}U_{\mathcal{L}_{1,2}}|\mathbf{0}\rangle$, where $U_{\mathcal{L}_{1,2}}$ is generated from a Hamiltonian with quadratic and linear products of terms from Jordan-Wigner Majoranas, we have at least efficient PIPO simulation {\color{edits} via a $(2n+1) \times (2n+1)$ covariance matrix.}
\end{theorem}

{\color{edits} Both results are a consequence from changes to which operators we use to define {\color{edits} the} covariance matrix. More importantly than the result itself is the consequence which comes from it; we can represent all stabilizer states via a covariance matrix. Stabilizer states are thus Gaussian, if we are open to Gaussianity not strictly being defined with respect to the Jordan-Wigner transform.
\\ \\
\noindent \textbf{Corollary 1}. \textit{For any stabilizer state $|\psi_{stab}\rangle$ there exist a set of mutually anticommuting Paulis $c_i \in \mathcal{L}_1^n$ such that $|\psi_{stab}\rangle$ is Gaussian.}}

\newpage

\section{Background}\label{sec:background}

\subsection{Fermions 
and Fermion-to-Qubit Encodings}

When discussing fermions, we use the language of second quantization. Fermionic models are then understood from products of creation and annihilation operators, which obey the anticommutation relations 
\begin{subequations}
    \begin{align}
        [a_i^{\dagger}, a_j^{\dagger} ]_+ &= [a_i, a_j ]_+ = 0 \\
        [a_i^{\dagger}, a_j ]_+ &= \delta_{ij}
    \end{align}
\end{subequations}
where $[a,b]_+ \equiv ab + ba$ denotes the anti-commutator. 
For a Fock basis state $|\textbf{f} \, \rangle = |f_1 f_2 f_3 ... f_n\rangle$ for $f_i \in \{0,1 \}$, we have that 
\begin{equation}
    \begin{aligned}
        a_j^{\dagger}|\textbf{f}\, \rangle &= \begin{cases}
    (-1)^{\pi_{j}}|{f_1   \dots , f_{j-1}  1 f_{j+1} \dots f_{n-1}}\rangle,& \text{if } f_j = 0\\
    0, & \text{if $f_j=1$}\, 
\end{cases} \\
    a_j|\textbf{f} \, \rangle &= \begin{cases}
    (-1)^{\pi_{j}}|{ f_1  \dots  f_{j-1} 0 f_{j+1} \dots f_{n-1}}\rangle,& \, \, \, \text{if } f_j = 1\\
    0,              & \, \, \, \text{if $f_j=0$}\, ,
\end{cases}
    \end{aligned}
\end{equation}
where $\pi_{j} \equiv {\sum_{i<j}f_i}$ over $\mathbb{F}_2$, the binary field over $\{0,1 \}$, is the parity of the occupied modes with index $i < j$. This phase factor comes from the anticommutation relations of fermionic operators. Thus, we can consider two calculations that need to be done when applying a fermionic operator to a state: an occupancy update, and a parity count calculation. \\
\indent With these operators, we can represent {\color{edits} arbitrary} fermionic Hamiltonians {\color{edits}. When we enforce parity and number  conservation, these Hamiltonians look like}
\begin{equation}
    H_{\text{M}} = \sum_{i,j}h_{ij} \ a_i^{\dagger}a_j + \sum_{i,j,k,l}h_{ijkl} \ a_i^{\dagger}a_j^{\dagger}a_k a_l { \; +\sum_{i,j,k,l,m,n} h_{ijklmn} \ a_i^{\dagger}a_j^{\dagger}a_k^{\dagger} a_l a_m a_n + ... \color{edits}.}
\end{equation}
{\color{edits} When} all $h_{ijkl} = {\color{edits} h_{ijklmn} = ...} =0$, $H_{\text{M}}$ can be efficiently and exactly diagonalized with a choice of a quasi-particle basis. {\color{edits} We can write}
\begin{equation}
    H_{\text{M}} = H_{\text{FF}} = \sum_i h'_i \  \Tilde{a}_i^{\dagger} \Tilde{a}_i {\color{edits},}
\end{equation}
where the set of $\Tilde{a}_i, \Tilde{a}_i^\dagger$ are linear combinations of the original fermionic creation and annihilation operators, and themselves also satisfy the fermionic anticommutation relations. We call the Hamiltonians made up of only quadratic products of fermionic operators \enquote{free} or \enquote{Gaussian}  fermionic Hamiltonians \cite{Surace_2022}. \\
\indent When working with fermions, it is often convenient to work with the Hermitian Majorana operators, defined as
% \begin{subequations}
%     \begin{align}
%     \gamma_{2k-1} &= a_k + a_k^{\dagger} {\color{edits},} \\
%     \gamma_{2 k} &= i( a_k- a^{\dagger}_k ) {\color{edits},} \\
%     [\gamma_i, \gamma_j]_+&= 2\delta_{ij} {\color{edits}.}
%     \end{align}
% \end{subequations}
\begin{align}
    \gamma_{2k-1} = a_k + a_k^{\dagger} {\color{edits},} \qquad
    \gamma_{2 k} = i( a_k- a^{\dagger}_k ) {\color{edits},} \qquad
    [\gamma_i, \gamma_j]_+&= 2\delta_{ij} {\color{edits}.}
    \end{align}
Majorana operators are often used in %the diagonalization of free fermion Hamiltonians. They also impose a more useful anticommutation structure for many uses in 
quantum information applications \cite{JozsaMiyake, chapman2023unified}. \\
\indent In order to simulate electronic systems on qubit devices, we need to represent the action of the fermionic operators on qubits. This is done using a fermion-to-qubit transform: for each fermionic operator, we define some qubit/spin operator such that the set of qubit/spin operators has the same anticommutation structure as the set of fermionic operators. In this regard, it is often more useful to use Majorana operators and choose a set of mutually anticommuting Pauli strings \cite{chien2020custom, Jiang_2020, HarrisonSerp}. \\

\noindent \textbf{Definition 1.} \textit{ {\color{edits} Let} $\{X, Y, Z, I \}$ {\color{edits} be} the collection of matrices \begin{equation}
    \begin{aligned}
        X & = \begin{pmatrix}
            0 & 1 \\ 1 & 0
        \end{pmatrix} 
        \; \; \; Y = \begin{pmatrix}
            0 & -i \\ i & 0
        \end{pmatrix} \\
        Z & = \begin{pmatrix}
            1 & 0 \\ 0 & -1
        \end{pmatrix} \; \; \; I = \begin{pmatrix}
            1 & 0 \\ 0 & 1
        \end{pmatrix}{\color{edits}.}
    \end{aligned}
\end{equation}The Pauli group is {\color{edits} the} set of operators \begin{equation}
    p_i = g_i \cdot p_{i_1} \otimes p_{i_2} \otimes ... \otimes p_{i_n} 
\end{equation}\\
for $p_{i_j} \in \{X, Y, Z, I \}$ and scalar phase $g_i \in \{1, i, -1, -i\}$, and group operation multiplication of elements.} \\ \\
Over $n$ qubits, the $4^n$ possible tensor product strings of the Pauli matrices, along with an arbitrary power of $i=\sqrt{-1}$, form the {\color{edits}elements of the} Pauli group $\mathcal{P}_n$. We refer to elements of the Pauli group as Pauli strings. For brevity, when dealing with Paulis strings we may omit the $\otimes$ and either write full strings of letters $X, Y, Z, I$ or specify only the non-trivial Paulis and the indices they act on, such as $X_1 X_3$. \\
\indent Note that all pairs of Pauli strings either commute or anti-commute. \\
\\
\noindent \textbf{Definition 2.} \textit{ 
A \emph{fermion-to-qubit encoding}
% or \emph{fermionic encoding} 
(or \emph{ transformation}) is an ordered set $\mathcal{L}_1^{(n)}$ of {\color{edits}$2n$} mutually anti-commuting Pauli strings:
} 
\begin{equation}
    \mathcal{L}_1^{(n)} \equiv \{ \ c_i \ | \ c_i \in \mathcal{P}_n, \ \{c_i, c_j\} = 2\delta_{ij} \ \}.
\end{equation}

We say that an encoding maps the $i^{\text{th}}$ Majorana operator to the $i^{\text{th}}$ element $\mathcal{L}_1 : \gamma_i \mapsto c_i$, and therefore occasionally use \emph{``an encoding's Majoranas''} as a synonym for the Pauli string elements of $\mathcal{L}_1^{(n)}$. We can denote the action of $\mathcal{L}_1^{(n)}$ on a fermionic Hamiltonian $H_F$ by $\mathcal{L}_1(H_F)$: the qubit operator given by making the substitution of $\gamma_i$ termwise by $c_i$. 
\\

\indent The choice of notation $\mathcal{L}_1^{(n)}$ will be discussed later in this paper. The most well known example of a fermion-to-qubit transform is the Jordan-Wigner transformation \cite{1928ZPhy...47..631J}. For $n$ qubits, this maps $2n$ Majoranas to $2n$ Pauli strings with (indices starting at 1)
\begin{align}\label{eq:JWMs}
    \gamma_{2i-1} &\xmapsto{\text{JW}} \left( \prod_{j=1}^{i-1} Z_j \right) X_i, \qquad
    \gamma_{2i} \xmapsto{\text{JW}} \left( \prod_{j=1}^{i-1} Z_j \right) Y_i .
\end{align}

With the Jordan-Wigner Majoranas defined, we can introduce the first circuit family we study in this paper. \\
\\
\noindent \textbf{Definition 3.} \textit{
An $n$-qubit unitary $U_{\text{MG}} \in U(2^n)$ is a \emph{matchgate circuit} that can be written as $U_{\text{MG}} = e^{-iH}$ for $H = \sum_{i,j}^{2n}h_{ij}c_i c_j$, written with the Jordan-Wigner encoding. Any matchgate circuit can be written as a circuit of $O(n^3)$ nearest neighbor gates $G(A,\,B)$ which may be written in the form 
\begin{align}{\label{eq:MG}}
G(A,\,B)=\begin{pmatrix} a_{11} & 0 & 0 & a_{12} \\
0 & b_{11} & b_{12} & 0 \\
0 & b_{21} & b_{22} & 0 \\
a_{21} & 0 & 0 & a_{22} \end{pmatrix}.
\end{align}
We define
\begin{align}
A\equiv \begin{pmatrix} a_{11} & a_{12} \\ a_{21} & a_{22} \end{pmatrix},\quad B\equiv \begin{pmatrix} b_{11} & b_{12} \\ b_{21} & b_{22} \end{pmatrix}{\color{edits},}
\end{align}
and require that the matrices $A$ and $B$ (both in $U(2)$) have the same determinant (i.e., $\det(A)=\det(B)$) \cite{valiant,TerhalDivincenzo, JozsaMiyake, knill2001fermionic}. 
}\\
\\

Though Jordan-Wigner is widely used, there are an infinite number of fermion to qubit encodings; the only requirement for a fermion-to-qubit encoding is that all operators mutually anticommute. Thus, we {\color{edits} can} conjugate {\color{edits} the} set of Jordan-Wigner Majoranas $\mathcal{L}_1^{\text{(JW)}} = \{c_1, c_2, c_3,... \, c_{2n} \}$ by any unitary $U$ and arrive at another set of mutually anticommuting operators (call it $\mathcal{S}^{(U)}$):
\begin{equation}
    \mathcal{S}^{(U)} = U \mathcal{L}_1^{\text{(JW)}} U^\dagger = \{ s_i = U c_i U^\dagger \ | \ c_i \in \mathcal{L}_1^{\text{(JW)}} \}{\color{edits}.}
\end{equation}
\indent Although $\{s_i, s_j\} = U \{c_i, c_j\} U^\dagger = \delta_{ij} $ for all $s_i$ and $\mathcal{S}^{(U)}$ constitutes a valid representation of the Majorana algebra, each $s_i$ would generally be a \emph{sum of Pauli strings} for arbitrary unitary $U$. If we instead restrict ourselves to the sets which we can reach by conjugating Jordan-Wigner by some Clifford $C$, we guarantee that we are always mapped to a set of individual mutually anti-commuting Pauli strings; indeed this property can be taken as a definition: \\
\\
\noindent \textbf{Definition 4.} \textit{
An $n$-qubit unitary $C \in U(2^n)$ is a \emph{Clifford operation} (or just \emph{a Clifford}) if it normalizes the Pauli group; that is, if conjugation by $C$ maps every Pauli string to another Pauli string: $ C p C^\dagger \in \mathcal{P}_n$ for all $p \in \mathcal{P}_n$. Clifford circuits are a finite group of unitary operators generated by
\begin{align*}
H = & \frac{1}{\sqrt{2}}\begin{pmatrix}
        1 & 1  \\
        1  & -1
    \end{pmatrix} {\color{edits},}\; \; \; S = \begin{pmatrix}
        1 & 0 \\ 0 & i 
    \end{pmatrix}{\color{edits},} \\
    &CNOT = \begin{pmatrix}
        1 & 0 & 0 & 0 \\
        0 & 1 & 0 & 0 \\
        0 & 0 & 0 & 1 \\
        0 & 0 & 1 & 0
    \end{pmatrix}{\color{edits},}
\end{align*} the Hadamard, Phase, and CNOT gates. 
}\\
\\
\indent By specifying any encoding $\mathcal{L}_1^{(n)}$ made up of individual Pauli strings, we also implicitly specify the Clifford which maps the $i^{\text{th}}$ Jordan-Wigner Majorana to the encoding's $i^{\text{th}}$ Majorana. For each choice of $\mathcal{L}_1^{(n)}$, there exists a Clifford $C$ mapping elements $d_i \in \mathcal{L}_1^{(n)}$ to elements $c_i \in \mathcal{L}_1^{\text{(JW)}}$ such that $d_i = C c_i C^\dagger$ for all $i$. This is a statement of the fact that \emph{all sets of mutually anticommuting Pauli strings are Clifford equivalent}. \\  \\
% (or sets of unitaries) 
% that conjugate $\mathcal{L}_1^{(n)}$ to and from the Jordan-Wigner encoding. Thus, by defining an encoding, we also define a Clifford circuit that takes us to that encoding from Jordan-Wigner.\\
\indent If $p \in \mathcal{P}_n$ is a Pauli string of length $n$ given by $p = p_{i_1} \otimes ... \otimes p_{i_n}$, then the \emph{locality} or \emph{Pauli weight} of $p$ is given by the number of non-identity letters $p_{i_j} \in \{X, Y, Z\}$. For quantum simulation purposes, we are often interested in choosing encodings whose Majoranas have low locality, as these can yield shorter quantum circuits in practice \cite{chien2020custom, HarrisonSerp}.\\
\indent One way to understand an encoding is to consider how information from Fock space is represented in qubit space. For the Jordan-Wigner encoding, the occupancy of one fermionic mode is stored only in a single qubit. While this has advantage for performing occupancy updates (which can be done in $O(1)$ cost) it does not for parity count operation (which is $O(n)$ cost for Jordan-Wigner). This cost of $O(n)$ for parity count summation is tied to the locality of {\color{edits} the} Jordan-Wigner Majoranas. \\
\indent For encodings with lower average locality, like the Bravyi-Kitaev encoding \cite{Setia_2018}, the occupancy of individual fermionic modes is stored across multiple different qubits. Compared to Jordan-Wigner, Bravyi-Kitaev increases the cost for occupancy updates to $O(\log_2(n))$ but also decreases the cost of parity count summation to $O(\log_2(n))$, achieving overall $\log$ scaling locality. A diagram for explaining how the occupancy information of the Bravyi-Kitaev encoding is stored across qubits is provided in Figure \ref{fig:FT}. 

\begin{figure}
    \centering
    \begin{tikzpicture}[->,>=stealth,auto,node distance=2cm,
  thick,main node/.style={circle,draw,font=\sffamily\Large\bfseries},state/.style={circle, draw, minimum size=2cm}]

  \node[main node] (1) {0};
  \node[main node] (2) [right of=1] {1};
  \node[main node] (3) [right of=2] {2};
  \node[main node,red] (4) [right of=3] {3};
  \node[main node] (5) [right of=4] {4};
  \node[main node] (6) [right of=5] {5};
  \node[main node] (7) [right of=6] {6};

  \node[line width = 0mm] (8) [below of=1, yshift=0.9cm]  {\LARGE $n_0$};
  \node[line width = 0mm] (9) [below of=2, yshift=0.9cm,xshift=0.25cm]  {\LARGE $n_0$};
  \node[line width = 0mm] (10) [below of=2, yshift=0.25cm]  {\LARGE $+n_1$};
\node[line width = 0mm] (11) [below of=3, yshift=0.9cm]  {\LARGE $n_2$};
  \node[line width = 0mm] (12) [below of=4, yshift=0.9cm,xshift=0.25cm]  {\LARGE $n_1$};
  \node[line width = 0mm] (13) [below of=4, yshift=0.25cm]  {\LARGE $+n_2$};
  \node[line width = 0mm] (14) [below of=13, yshift=1.35cm]  {\LARGE $\color{red}+n_3$};
    \node[line width = 0mm] (15) [below of=5, yshift=0.9cm]  {\LARGE $n_4$};
  \node[line width = 0mm] (16) [below of=6, yshift=0.9cm,xshift=0.25cm]  {\LARGE $n_4$};
  \node[line width = 0mm] (17) [below of=6, yshift=0.25cm]  {\LARGE $+n_5$};

  \node[line width = 0mm] (18) [below of=7, yshift=0.9cm,xshift=0.25cm]  {\LARGE $\color{red}n_3$};
  \node[line width = 0mm] (19) [below of=7, yshift=0.25cm]  {\LARGE $+n_5$};
  \node[line width = 0mm] (20) [below of=19, yshift=1.35cm]  {\LARGE $+n_6$};

  \path[every node/.style={font=\sffamily\small}]
    (2) edge[bend right] node [left] {} (1)
    (4) edge[bend right] node [left] {} (2)
    (4) edge[bend right] node [left] {} (3)
    (7) edge[bend right] node [left] {} (6)
    (6) edge[bend right] node [left] {} (5)
    (7) edge[bend right,red] node [left] {} (4);
    
\end{tikzpicture}
    \caption{Occupancy graph for Bravyi-Kitaev encoding, which corresponds to the Fenwick tree data structure. Each node stores the occupancy information of itself, as well as occupancy of any of its children. While compared to Jordan-Wigner (where nodes only store their own occupancy) occupancy needs to be updated in multiple sites, parity information/parity count can be calculated without needing to query all nodes, and thus overall locality cost is $O(\log_2(n))$ \cite{Setia_2018}. As a specific example, updating the occupancy of node $3$ (shown in red) requires an update to the occupancy of node $6$ as well.} 
    \label{fig:FT}
\end{figure}

\subsection{Definitions used in Proofs of Simulability}

Here we clarify important definitions we use throughout the paper, which will be used to describe simulability classes. \\

\noindent \textbf{Definition 5.} \textit{
A \emph{Computational basis state Input (CI)} is a single bitstring qubit state in the $Z$ basis, such as $|\psi\rangle = |0\rangle_1 \otimes |0\rangle_2 \otimes ... \otimes |0\rangle_n \equiv |\mathbf{0}\rangle$
}.\\

\noindent \textbf{Definition 6.} \textit{
A \emph{Product-state Input (PI)} is a state $|\psi_{\textnormal{prod}}\rangle = \bigotimes_{i=1}^n |\psi _ i\rangle$, where $|\psi_i\rangle$ is an arbitrary single qubit state. That is, $|\psi_{\textnormal{prod}}\rangle = \bigotimes_{i=1}^n (a_i|0\rangle + b_i|1\rangle)$ for $a_i, b_i \in \mathcal{C}$ and {\color{edits} with} $|a_i|^2 + |b_i|^2 = 1$ $\forall \; i \in [  1, n]$.} \\

\noindent Here $[1,n]$ is the set $\{1,2,...,n\}$.

\indent We are interested in what outputs it is possible to efficiently compute for a given circuit, with respect to these inputs, as well as for arbitrary input states. Therefore, we clarify the following sets of observables. \\

\noindent \textbf{Definition 7.} \textit{
\emph{Bitstring outputs (BO)}, for a given {\color{edits} $n$-qubit }unitary $U$ and initial qubit state $|x\rangle$, are the marginal measurement probabilities $p(y^*)$ of finding a particular subset of $k$-many qubits to be the bitstring $y^* \in \{0,1\}^{ k}$. (Without loss of generality we can assume $y^*$ is a bitstring over the first $k$ qubits). These are given by
\begin{equation}
    p(y^*) = \langle x | U^{\dagger} |y^*\rangle\langle y^*| U | x\rangle,
\end{equation}
where for $k < n$ we have the projector explicitly as
% \begin{equation}
$
    |y^*\rangle \langle y^*| = \left(\bigotimes_{i=1}^{k} |y_i\rangle\langle y_i| \right) \otimes I_{n-k}
% \end{equation}
$.} \\

\indent An algorithm is said to be \emph{capable of bitstring outputs} with respect to {\color{edits} $n$-qubit unitary} $U$ and $|x\rangle$ if the cost of calculating $p(y^*)$ for any $k\in [1, n]$ and any $y^* \in \{0, 1\}^{k}$ scales at most polynomial in both $n$ and $k$.

\newpage

\noindent \textbf{Definition 8.} \textit{
\emph{Restricted bitstring outputs (denoted bO)} are the bitstring outputs for {\color{edits} unitary} $U$ and {\color{edits}input state} $|x\rangle$ over only $n$-qubit bitstrings: $p(y^*) = |\langle y^* | U | x \rangle|^2 $ for $y^* \in \{0, 1\}^{n}$. 
}\\

An algorithm is said to be \emph{capable of restricted bitstring outputs} with respect to $U$ and $|x\rangle$ if the cost of calculating $p(y^*)$ for any $y^* \in \{0, 1\}^{n}$ scales at most polynomially in $n$. Such an algorithm is not necessarily capable of computing all marginal probabilities, but is always able to compute the measurement probability of a particular full bitstring over the register of $n$ qubits. \\

\noindent \textbf{Definition 9.} \textit{
\emph{Pauli expectation Outputs (PO)}, for a given unitary $U$ and input state $|x\rangle$, are the expectation values $\langle p \rangle = \langle x | U^\dagger p U | x\rangle$ for $p \in \mathcal{P}_n$.
}\\

\indent An algorithm is said to be \emph{capable of Pauli expectation outputs} with respect to $U$ and $|x\rangle$ if the cost of calculating $\langle p \rangle$ for any $p \in \mathcal{P}_n$ scales at most polynomially in $n$. \\

\noindent \textbf{Definition 10.} \textit{
\emph{Restricted Pauli expectation Outputs (denoted pO)} for a given $U$ and $|x\rangle$, together with an encoding $\mathcal{L}_1 = \{ c_i \}$, are the Pauli expectation outputs restricted to the Pauli strings which can be written as products of some constant number $d = O(1)$ of the encoding's Majoranas  $\mathcal{L}_1^{(n)}$. We can write this set of restricted Pauli strings $\mathcal{P}^{(r)}$ as the products of all selections $\sigma \subset \{1, 2, ... \, 2n\}$ of at most $d$-many of the elements $c_i \in \mathcal{L}_1^{(n)}$: 
\begin{equation}
    \mathcal{P}_n^{(r)} = \left\{ p^{(r)} \ \Bigg| \ p^{(r)} = \prod_{i \in \sigma } c_{i}, \ \sigma \subset \{1, ... \ 2n\}, \ |\sigma| \leq d \ \right\}{\color{edits}.}
    % \ \Big| \ d = O(1)
\end{equation}
Then the restricted Pauli expectation outputs are the Pauli expectation values $\langle x | U^\dagger p^{(r)} U | x \rangle $ for all of the restricted Pauli strings $p^{(r)} \in \mathcal{P}_n^{(r)}$.} \\
\indent Note that for increasing $n$, the number of such restricted {\color{edits} Pauli strings} scales asymptotically as $O(n^d)$. These outputs constitute a polynomially large subset of the expectation values of Pauli strings over $n$ qubits. 
\\

An algorithm is said to be \emph{capable of restricted Pauli expectation values} with respect to some unitary $U$ and input state $|x\rangle$, together with an encoding $\mathcal{L}_1^{(n)}$, if the cost of computing $\langle p^{(r)} \rangle$ for any restricted Pauli string $p^{(r)} \in \mathcal{P}_n^{(r)}$ scales polynomially with $n$. Such algorithms (which are not also capable of arbitrary PO) rely on the structure of the constant product of an encoding's Majoranas to compute these expectation values, as we will see. \\

For clarity with the literature \cite{JoszaCliffC, MatchwOutside}, we highlight the notions of \emph{strong} versus \emph{weak} simulability of a quantum circuit. A \textit{strong simulation} of quantum circuit $U$ on input $|x\rangle$ is defined to be a classical algorithm which is able to compute any of the marginal probabilities in the bitstring output (BO) setting in polynomial time i.e. a classical algorithm which is capable of bitstring outputs with respect to $U$ and $|x\rangle$. 
% ; i.e., we have efficient computation of any desired output probability. 
In contrast, a \emph{weak simulation} is a classical algorithm which can produce \emph{samples} of bitstrings of length $k$ according to the corresponding $k$-qubit marginal measurement probability distribution of $U|x\rangle$.
In this paper, we focus on the \textit{strong simulation} of circuits{\color{edits}, though it is important to note that \textit{strong simulation} allows for efficient \textit{weak simulation}.}
% For bitstring outputs (MO) and Pauli expectation value outputs (PO), we discuss \emph{strong} simulability,
% though for Pauli expectation values it's only strong over single qubit outputs 
\cite{JoszaCliffC, MatchwOutside, terhal2004adaptivequantumcomputationconstant}.\\

\noindent \textbf{Definition 11.} \textit{
A family of quantum circuits $\mathcal{C}$ is said to be \emph{in the simulability class} determined by a set of states $S_{\textnormal{IN}}$ and output observables $O_{\textnormal{OUT}}$ {\color{edits}(either bitstring output probabilities or Pauli observables)} if for all $C \in \mathcal{C}$ and all $|x \rangle \in \mathcal{S}_{IN}$, there exists a classical algorithm which is capable of computing all values $v \in O_{\textnormal{OUT}}$ in polynomial time. 
}\\

\indent To describe different classes of inputs and outputs we consider, we refer to the acronym chart in figure \ref{fig:aTable}. These acronyms, while similar to acronyms used in previous matchgate literature \cite{Brod_2016, mocherla2024extending}, are modified to better fit this paper. For example, the acronym ``CIBO'' refers to the ability to efficiently calculate any marginal bitstring probability after applying the circuit in question to any computational basis state. \\

\begin{figure*}\label{fig:aTable}
    \textbf{Inputs} \hspace{5cm} \textbf{Outputs} \vspace{1cm}
    \centering
    \begin{tabular}{ |c||c|c|c|c| } 
    \hline
     & Bitstring & Pauli & Restricted Bitstring & Restricted Pauli \\
    \hline
    \hline
    Computational basis state (CI) & CIBO & CIPO & CIbO & CIpO\\ 
    \hline
    Product state (PI) & PIBO & PIPO & PIbO & PIpO\\ 
    \hline
    \end{tabular}
    \caption{Table of acronyms used throughout the paper (while defined for clarity, no family of circuits we consider in this paper belong to the classes described by  CIbO nor PIBO). Note certain %clear 
    inclusions exist: as an example,
    CIbO $\subset$ CIBO $\subset$ PIBO.}
\end{figure*}

%We will define the following acronyms.
%\begin{itemize}
%    \item PIBO (CIBO): product state input (computational basis state input), many output
%    \item CIBO: computational basis state input, restricted many qubit output
%    \item PIPO (CIPO): product state input (computational basis state input), Pauli expectation value output
%    \item PIpO: product state input, restricted Pauli expectation value output
%\end{itemize}
 \subsection{Previous work on the Simulability of Matchgates and Cliffords}

\subsubsection{Cliffords}

When we apply a Clifford, we can track the evolution of the state by understanding how the Clifford conjugates all of {\color{edits} the} stabilizers $Z_i$. Classically simulating these circuits is most often done with a Clifford tableaux, a $2n \times 2n+1$ matrix over $\mathbb{F}_2$ which stores Pauli and phase information \cite{Aaronson_2004}. \\
\indent Instead of %We can also consider, rather than on 
computational basis state inputs, we can also consider Cliffords acting on product state inputs. For expectation values of Pauli strings, we can simulate expectation values over Cliffords acting on product states in polynomial time. The expectation value
\begin{equation}
    \langle \psi_{\text{prod}}| C^{\dagger} p C| \psi_{\text{prod}} \rangle = \langle \psi_{\text{prod}}| p' |  \psi_{\text{prod}}\rangle
\end{equation}
is the expectation of a product operator ($U_1 \otimes U_2 \otimes ... U_n$) on a product state, which can be calculated in polynomial time \cite{CliffonProd}. \\
\indent On computational basis state inputs we can simulate both probability marginals and Pauli expectation values of Clifford circuits; on product state inputs, we can only simulate Pauli expectation values in general \cite{PhysRevLett.123.170502}.
\subsubsection{Matchgates}
Originally proven to be simulable in the context of the perfect matching problem on planar graphs \cite{valiant}, small changes to the structure of matchgates such as relaxation of the determinant condition or the allowance of next-nearest neighbor gates \cite{Brod_2012, BrodXY, JozsaMiyake, Bravyi_2002} are sufficient to realize universal quantum computation. In terms of a single gate to add to {\color{edits}the set of} matchgates to recover universal computation, the SWAP gate is discussed most often. However, any two qubit parity preserving gate without the determinant condition can also be used as a universal resource \cite{TerhalDivincenzo, JozsaMiyake, Brod_2016}. \\\\
\indent  Although simulability in matchgate circuits can be understood via the connection to the perfect matching problem, a deeper physical intuition behind their simulability can be found from their connection to free fermions \cite{knill2001fermionic, TerhalDivincenzo}. \\
\indent Matchgates specifically come from the exponentiation of two-qubit nearest neighbor free (quadratic) fermionic Hamiltonians, which may be written with using some encoding's Majorana and real coefficients as
\begin{equation}
    \begin{aligned}
        H_{F} = i ( & \alpha_0 c_{2k-1} c_{2k+2} - \alpha_1 c_{2k}c_{2k+1} \\
         + & \beta_1 c_{2k-1}c_{2k+1} - \beta_2 c_{2k}c_{2k+2} \\
         - & \delta_1 c_{2k-1}c_{2k} - \delta_2 c_{2k+1}c_{2k+2}).
    \end{aligned}
\end{equation}
Via the Jordan-Wigner transform, we can define an equivalent spin Hamiltonian which produces matchgates, 
\begin{equation}
\begin{aligned}
    H_{S} = (&\alpha_0 Y_k Y_{k+1} + \alpha_1 X_k X_{k+1} \\
    + &\beta_1 Y_k X_{k+1} + \beta_2 X_k Y_{k+1} \\
    + &\delta_1 Z_k + \delta_2 Z_{k+1}).
\end{aligned}\label{eq:JWH}
\end{equation}
As shown in \cite{Jozsa_2009}, the exponentiation of {\it{any}} free fermion Hamiltonian can be written as the product of $O(n^3)$ matchgates. \\
\indent The key fact that allows for simulability of free fermion dynamics relies on the conjugation of a fermionic operator by a unitary generated from any free fermion Hamiltonian. We will review this calculation for Majoranas, though a similar calculation can be done for creation/annihilation operators. \\
\indent Let our free fermion Hamiltonian be written as $H_F = i\sum_{i,j}h_{i,j}c_i c_j$ for $h_{i,j} \in \mathcal{R}$, and let $U(t) = e^{itH_F}$. We will ask about the time translation of a fermionic operator $c_i(t) = U^{\dagger}(t) c_i(0) U(t)$. We can solve for the time translation by solving the differential equation
\begin{align*}
    \frac{d c_u(t)}{d t} &= i \left[H_F, c_u\right] \\
	&= -\left[\sum_{i,j=1}^{2n} h_{ij}c_i c_j, c_u\right] \\
	&= \sum_{v=1}^{2n} 4h_{uv}c_v(t)
\end{align*}
with solution
\begin{align}\label{eq:compact}
    c_u &= \sum_{v=1}^{2n} R_{uv}c_v \\
	R_{uv} &= e^{4h} \in O(2n){\color{edits}.}
\end{align}
We thus can see that the time translation of fermionic operators by free fermion Hamiltonians can be calculated in polynomial time. \\
\indent We can now go through the proofs and discussion of different simulable inputs and outputs for matchgate circuits. 

\subsubsection{CIBO}

Computational input/bitstring output (CIBO) simulation is defined by Terhal and Divincenzo \cite{TerhalDivincenzo}. We will go through a sketch of their proof of simulability with the use of creation and annihilation operators, but for more context and for a proof which uses Majorana operators we refer to the full paper \cite{TerhalDivincenzo}. \\
\indent We will let {\color{edits} the} input state be the state $|\mathbf{0}\rangle$ (without loss of generality to any other computational basis state), and ask about the probability of measuring some $k$-bit substring of outputs to be in bitstring $y^*$ after free fermion evolution. We write 
\begin{equation*}
    p(y^*) = \langle \mathbf{0} | U_{\text{MG}}^{\dagger} |y^* \rangle \langle y^* | U_{\text{MG}} |\mathbf{0} \rangle{\color{edits}.}
\end{equation*}
We can write the projector of $|y^*\rangle\langle y^*| $ as a string of fermionic operators, where index $i$ being $0$ corresponds to string $a_ia_i^{\dagger}$ and index $j$ being $1$ corresponds to string $a_j^{\dagger}a_j$. Thus, defining output bitstring $y^* = 01...0$ without loss of generality we can rewrite the projector in {\color{edits} the} equation with
\begin{equation}
    p(y^*) = \langle \mathbf{0} | U_{\text{MG}}^{\dagger} a_{i_1}a_{i_1}^{\dagger}a_{i_2}^{\dagger}a_{i_2}...a_{i_k}a_{i_k}^{\dagger} U_{\text{MG}} |\mathbf{0} \rangle{\color{edits}.}
\end{equation} 
{\color{edits} Inserting} the identity $U_{\text{MG}}U^{\dagger}_{\text{MG}}$ in front of every fermionic operator, we can write and then use conjugation rules to find
\begin{equation}
    p(y^*) = \sum_{a,b,...,k}T_{i_1, a}T_{i_2, b}...T_{i_k, k} \langle \mathbf{0}|a_a a_b ... a_k|\mathbf{0}\rangle{\color{edits},}
\end{equation}
{\color{edits} for $T$ the rotation matrix defined by deriving equation \hyperref[eq:compact]{19} with creation/annihilation operators instead of Majoranas}.
While this summation may appear exponentially scaling, due to Wick's Theorem \cite{PhysRev.80.268} the expectation values can be simplified greatly. The entire calculation written as
\begin{equation}
    p(y^*) = \text{Pf}(M){\color{edits},}
\end{equation}
the Pfaffian of some antisymmetric matrix $M=(m_{ij})$. For a $2n \times 2n$ antisymmetric matrix, the Pfaffian can be written as 
\begin{align}
    \text{Pf}(M) = \frac{1}{2^n n!} \sum_{\sigma \in S_{2n}} \text{sgn}(\sigma)\prod_{i=1}^n m_{\sigma(2i-1)\sigma(2i)} = \sqrt{\text{det}(M)}
\end{align}
for $S_{2n}$ the symmetric group of degree $2n$. The sign of permutation $\sigma$ is the parity of the number of transpositions used to describe $\sigma$. As {\color{edits} determinants are easy to compute}, the Pfaffian {\color{edits} of any antisymmetric matrix} is also {\color{edits} easy} to compute. 

\subsubsection{PIpO}
Product state input/restricted Pauli output (PIpO) simulability is defined by Jozsa and Miyake \cite{JozsaMiyake}. Following Jozsa and Miyake, we focus on the simulation of the Pauli $Z_i$ expectation value. \\
\indent Let {\color{edits} the} task be defined as computing
\begin{equation}
    \langle Z_i \rangle = \langle \psi_{\text{prod}} | U_{\text{MG}}^{\dagger} Z_i U_{\text{MG}} | \psi_{\text{prod}} \rangle {\color{edits}.}
\end{equation}
Using the Jordan-Wigner encoding, we can write $Z_i = c_{2i-1}c_{2i}$, and by inserting a the identity $Z_i = c_{2i-1}U_{\text{MG}}U_{\text{MG}}^{\dagger}c_{2i}$. Plugging this in for $Z_i$ and conjugating {\color{edits} the} Majorana operators accordingly, we can write 
\begin{equation}
     \langle Z_i \rangle = \sum_{j,k}R_{j,2i-1}R_{k,2i}\langle \psi_{\text{prod}} | c_j c_k | \psi_{\text{prod}} \rangle {\color{edits}.}
\end{equation}
The summation is of order $n^2$, and the evaluation of $\langle \psi_{\text{prod}} | c_j c_k | \psi_{\text{prod}} \rangle$ can be done in polynomial time with the  acknowledgement that $c_j c_k$ is some Pauli string for all $j$ and $k$. \\
\indent For any Pauli string which can be written as a constant scaling number of Majoranas, $p = \prod_{i=1}^{d} c_{j_i}$ for $d \in O(1)$, the expectation value 
\begin{equation}
    \langle Z_i \rangle = \langle \psi_{\text{prod}} | U_{\text{MG}}^{\dagger} p U_{\text{MG}} | \psi_{\text{prod}} \rangle
\end{equation}
can be calculated in polynomial time, with near identical proof as seen for $Z_i$. 

\subsubsection{PIBO and the Hadamard Gadget}\label{app:hdevice}

While the proofs of simulability for CIBO and PIpO seem separate, Daniel Brod was able to unite the two into PIBO, allowing for many qubit output simulation on product state inputs \cite{Brod_2016}. The key observation %that Daniel Brod makes 
is with only a single ancilla qubit in the $|+\rangle$ state, product states can be made from free fermion time dynamics. Brod defines the 'Hadamard device', which relies on {\color{edits} two} key identities {\color{edits}:}
\begin{align*}
    G(H,H)|\psi_{\text{prod}}\rangle|+\rangle &= H|\psi_{\text{prod}}\rangle |+\rangle {\color{edits},} \\
    G(Z,X)|\psi_{\text{prod}}\rangle|0\rangle &= |0\rangle|\psi_{\text{prod}}\rangle {\color{edits},} \\
\end{align*}
where $G(Z,X) = fSWAP$. With this, we can construct product states as follows. 
\begin{enumerate}
    \item Apply repeated applications of $G(H,H)$ on the qubit next to the ancilla and the ancilla, interspersed with $R_Z$ gates on the qubit next to ancilla. 
    \item After enough repeated applications, the qubit state next to the ancilla can be in any arbitrary product state.
    \item fSWAP the product state up the chain of qubits.
    \item Repeat this process until all qubits besides the ancilla are in arbitrary product states.
\end{enumerate}

Thus, we can see that for {\color{edits} generating} the product state input the only non-Gaussian element needed is the ancilla in the $|+\rangle$ state. We can then write, for measuring some bitstring output $y^*$ on some product state input $|\psi_{prod}\rangle = U_{\text{PS}}(1+a_{n+1}^{\dagger})|\mathbf{0} \rangle $ ($U_{PS}$ a matchgate circuit) as 
\begin{align*}
       p(y^*) &= 
        \frac{1}{2} \langle \mathbf{0} | (1+a_{n+1})U_{\text{PS}}^{\dagger}U_{\text{MG}}^{\dagger} |y^*\rangle \langle y^*| U_{\text{MG}}U_{\text{PS}}(1+a_{n+1}^{\dagger})|\mathbf{0} \rangle \\
        &= \frac{1}{2}(\text{Pf}(M_1) + \text{Pf}(M_2)) {\color{edits},}
\end{align*}
where the output is the sum of two Pfaffians, rather than a single Pfaffian.

\subsection{Product States}

Before we continue, we will discuss ways to make product states, without the use of measurement. We will highlight two constructions. \\
\indent  The first, and more conventional way to think about product states, is to have access to local rotations on every qubit in {\color{edits} the} system. While the Hamiltonian generators of the necessary unitaries are not unique, access to arbitrary $X$ and $Z$ rotations is sufficient. A common set of generators is $\{X_1, X_2, ... ,X_n, Z_1, Z_2, ..., Z_n \}$. \\
\indent However, there is another way we can think to make arbitrary product states, even without access to local non-commuting rotations on each qubit. We could instead require that for \textit{some} qubit $k$ (or multiple qubits), that we have access to $\{X_k, Z_k\}$, and then the ability to do a SWAP-like operation from qubit $k$ to any other qubit in the system. We can define {\color{edits}a} SWAP-like element as 
\begin{align*}
    \begin{pmatrix}
        q_1 & 0 & 0 & 0 \\
        0 & 0 & q_2 & 0 \\
        0 & q_3 & 0 & 0 \\
        0 & 0 & 0 & q_4
    \end{pmatrix}{\color{edits},}
\end{align*}
where each {\color{edits}$q_i = e^{i\theta}, \theta \in [0,2\pi]$}. A gate of this form is generated by the Hamiltonian $H = q' (X_iX_j + Y_iY_j) + \theta_1 Z_iZ_j + \theta_2 (Z_i + Z_j) + \theta_3 I$. While the angles $\theta_i \in \mathcal{R}$ are free, {\color{edits} we fix} $q' = \frac{2N+1}{4}\pi$ for $N\in \mathcal{Z}$. Rather than having local rotations, with access to local rotations on at least one qubit and access to the $XX$ and $YY$ generators, we can still make product states. \\
\indent {\color{edits}We} note that depending on the construction we use for product states, the SWAP-like operation can be an entangling gate that does not create entanglement on certain product states. As an example, fSWAP ($q_1 = q_2 = q_3 = -q_4 = 1$) has the power to entangle arbitrary product states, though {\color{edits} it }do{\color{edits}es} not create entanglement on two qubit states of the form $| \psi_i \rangle | 0\rangle$. \\
\indent We can observe that this is how product states are made for matchgates. As discussed above, Daniel Brod's Hadamard device \cite{Brod_2016}, and $Z$ rotation gates $R_Z = e^{-i \theta Z}$, gives access to $X$ and $Z$ rotations on a single qubit. The product state made is then swapped up the chain by Gaussian operations {\color{edits}(i.e, unitaries generated by free fermion mappable Hamiltonians)}, because both $XX$ and $YY$ can be mapped to quadratic products of Majoranas. \\
\indent However, while the Hadamard device utilizes a seemingly non-Gaussian resource to generate product states, we argue that for matchgates, product states can be made from purely Gaussian dynamics. The realization comes from recognizing that rather than the Hadamard device, we could instead ask to have access to the the generator $X_1$ {\color{edits} for our unitary dynamics}. This is granted by having access to the dynamics of single Majorana operators, which for Jordan-Wigner includes $X_1$.\\
\indent We present two Lemmas from previous work and sketch their proofs to highlight methods we refer to throughout our paper. Both Lemmas show how to map Hamiltonians of quadratic and linear terms of Majoranas to solely quadratic Hamiltonians, though they use different techniques. {\color{edits} In the first mapping, we will exactly recover product states at the cost of a mapping to a physical system. For the second mapping, we will recover product states to an isomorphism from non-product states over $n+1$ qubits to product states over $n$ qubits. In either, \enquote{product} states can be made with purely Gaussian dynamics over {\color{edits} the} adjusted set of Majorana operators, and thus are Gaussian.}\\

%Our first two corollaries are ...

\noindent \textbf{Lemma 1} \textit{\label{cor:ExtL}
   \emph{(Theorem 3 from Jozsa, Miyake, Strelchuk \cite{jozsa2015jordanwigner})} Let $H_F$ be a fermionic Hamiltonian $H_F = i\sum_{j,k=1}^{2n} h_{jk}\gamma_j \gamma_k + \sum_{j=1}^{2n} b_j \gamma_j$, with linear and quadratic products of $2n$ Majoranas. Then there exists a purely quadratic fermionic Hamiltonian $H_F' = i\sum_{j,k=0}^{2n} h_{jk}'\gamma_j' \gamma_k'$ with operators $[\gamma'_j, \gamma'_k]_+ = 2\delta_{jk}$
    % $2n+1$ 
    such that $H_F' = H_F$.  
}
\begin{proof}
    We first note that $h$ must be real and antisymmetric, and $b_i$ purely real, to guarantee Hamiltonian $H_F$ is Hermitian. \\
\indent For a set of $2n$ mutually anticommuting Pauli strings/fermionic operators (acting on $n$ qubits/modes), it is always possible to construct at least one more operator which also anticommutes with all existing elements \cite{sarkar2019setscommutinganticommutingpaulis}. 
By appending each Pauli string in a mutually anticommuting set with either $I$ or $Z$, we can guarantee the existence of another Pauli string on $n+1$ qubits (indeed several choices for this are described in \cite{anticomm}). For this proof, we can assume without loss of generality that we have added {\color{edits} the} total parity operator to {\color{edits} the} set of Majoranas, $\gamma_0 = \prod_{i=1}^{2n} \gamma_i$. We can then define a new set of $2n+1$ operators $\gamma_i'$, where the first Majorana is unchanged $\gamma_0' = \gamma_0$ and the remaining $2n$ are defined as $\gamma_i' = i \gamma_i \gamma_0$. It is easy to show that these primed operators satisfy the Majorana anticommutation relations. We can then write, 
\begin{align*}
    H_F' &= i\sum_{i,j=0}^{2n} h_{ij}'\gamma_i' \gamma_j'  \\
    &= i\sum_{j=1}^{2n} h_{0,j} \,  \gamma_0 i \gamma_j \gamma_0 - i\sum_{i,j=1}^{2n} h_{ij} \, \gamma_i \gamma_0 \gamma_j \gamma_0 \\
    &= i\sum_{i,j=1}^{2n} h_{ij} \, \gamma_i \gamma_j + \sum_{i=0}^{2n} b_i \, \gamma_i = H_F {\color{edits},}
\end{align*}
where $h_{0,j} = -b_j/2$ % \frac{b_j}{2}$ 
and $h_{j,0} = b_j/2$ \cite{jozsa2015jordanwigner}. Thus, by adding a single additional Majorana, we can write a Hamiltonian which has single and quadratic fermionic terms as a purely quadratic fermionic Hamiltonian. \\
\end{proof}
 
\indent We can write another Lemma that specifically utilizes extending {\color{edits} the} system of $n$ qubits to $n+1$ qubits. Before continuing, we will borrow notation from Knill \cite{knill2001fermionic}, defining, in addition to $\mathcal{L}_{1}^{(n)}$
\begin{itemize}
    \item $\mathcal{L}_{2}^{(n)}$: The set of quadratic products of an encoding's Majoranas $\{ c_i c_j | c_k \in \mathcal{L}_{1}^{(n)} \}$, excluding identical terms $c_k^2 = 1$.
    \item $\mathcal{L}_{1,2}^{(n)}$: The set of linear and quadratic product of an encoding's Majoranas $\mathcal{L}_{1}^{(n)} \cup \mathcal{L}_{2}^{(n)}$. 
\end{itemize}
% If we need to indicate that a set of operators defined over $k$ qubits rather than $n$, we will write $\mathcal{L}^{(k)}$.

\noindent \textbf{Lemma 2} \textit{
    (\emph{Section 6, Knill \cite{knill2001fermionic}}) Let $\mathcal{L}_{{1,2}}^{(n)}$ be the linear and quadratic products of Jordan-Wigner Majoranas on $n$ qubits. There exist an isomorphism from $\mathcal{L}_{{1,2}}^{(n)}$ to a subset of $\mathcal{L}_{2}^{(n+1)}$, the quadratic products of the Jordan-Wigner encoding over $n+1$ qubits. 
}
\begin{proof}
For the Jordan-Wigner over $n$ qubits, {\color{edits} defined by equation \ref{eq:JWMs}, taking their products gives}
\begin{align}
    \LL_{2}^{(n)} = \left\{ 
    U_{\text{I}}^{(a)} \left(\prod_{k=i+1}^{j-1} Z_k \right) U_j^{(b)} \, \Bigg| \, \begin{array}{l}
    U^{(a)}, U^{(b)} \in \{X, Y\} \\
    1 \leq i \leq j \leq n
    \end{array}
    \right\}\label{eq:LL2} {\color{edits}.}
\end{align}
Note that for $i = j$ above we recover the elements $Z_i$  (up to phases) for $1 \leq i \leq n$ and $U^{(a)} \neq U^{(b)}$.
    %For Jordan-Wigner on $n$ qubits, the set $\mathcal{L}_{{1,2}}^{(n)}$ consist of operators $X,Y,ZX,ZY, ZZX, ZZY, ...$, $Z_i$ for all $i$, and Pauli strings of the form $U_{\text{I}}Z_{i+1}Z_{i+2}...Z_{k-1}U_k$, for $U_{\text{I}}$ and $U_k$ in $\{X, Y\}$. 
    
    We define an isomorphism as follows: for each element $c_i$ in $\mathcal{L}_{{1,2}}^{(n)}$, let
    \begin{equation*}
        c_i^* = \begin{cases}
    I_0\otimes c_i,& \text{if $c_i \in \mathcal{L}_{{2}}^{(n)}$}\\
    X_0 \otimes c_i, & \text{if $c_i \in \mathcal{L}_{{1}}^{(n)}$}\, {\color{edits}.}
\end{cases}
    \end{equation*}
        Recall that all terms in $\mathcal{L}_1^{(n)}$ are of the form $c_j=  \prod_{i=1}^{j-1} Z_i U_j$ for some $1 \leq j \leq n$ and $U  \in \{X,Y \}$. Thus, $X \otimes c_j$ for $c_j\in \mathcal{L}_1^{(n)}$ is of the form $ X_0 (\prod_{k=1}^{j-1} Z_k) U_j, \; U \in \{X, Y\}$. Now, all $c^*_j$ are of the form 
    given in \eqref{eq:LL2} for $n+1$ qubits. %Every element $c_i^*$ is contained in $\mathcal{L}_{2}^{(n+1)}$. 
    %It is then easy to show 
    This completes the proof that the map defined by $c_i \mapsto c_i^*$ is an isomorphism between $\mathcal{L}_{{1,2}}^{(n)}$ and the subset of operators in $\mathcal{L}_{2}^{(n+1)}$ whose Pauli strings begin with either $I$ or $X$.
\end{proof}

 \indent With both of these proofs, we have access to dynamics over single Majoranas due to their mapping to a free fermion Hamiltonian. For Jordan-Wigner, the ability to map the dynamics of terms in $\mathcal{L}_{1,2}^{(n)}$ to purely quadratic models indicates that a product state, in the context of Jordan-Wigner, is a Gaussian resource. 

\section{Proofs of Complexity of Clifford Conjugated Matchgates}\label{sec:structure}

{\color{edits} We now begin to go through a hierarchy of Clifford conjugated matchgates, and their simulability (or lack thereof). As we introduce more Cliffords and allow for  fermionic information to be spread across qubits in different ways, we will accordingly lose access to simulable outputs. We diagram our results in Figure \hyperref[fig:pyra]{3}.}

% \begin{figure*}
%   \centering
%   \label{fig:pyra}
%   \includegraphics[width=\textwidth]{New_imag.png}
%   \caption{A visual guide to the hierarchy of Clifford conjugated matchgate circuits we are going to consider. As we go up the hierarchy, {\color{edits} we allow for the action of Cliffords to act more freely. Inversely, as we allow for more Cliffords to act we lose versions of simulability. We begin with circuits for which we can compute all bitstring outputs and marginals on various inputs. As we go up, we lose inputs/outputs for which marginals are simulable, and eventually lose marginals altogether. We end} with circuits for which only a polynomial number of Pauli expectation values are simulable, {\color{edits} and for which we conjecture sampling is hard}. {\color{edits} $S$ represents SWAP circuits,} $CZ$ represents control-$Z$ circuits, $C_{\text{P}}$ arbitrary Clifford permutations, and $C$ arbitrary Cliffords. }
% \end{figure*}

\begin{figure*}
  \centering
  \label{fig:pyra}
  % \begin{center}
\scalebox{0.75}{
\begin{tikzpicture}
\node[trapezium,
    draw, text centered, fill = white,  trapezium left angle=45, trapezium right angle=45,  text width = 309, text height = 9] (t) at (0, 2.75) {Conjectured Hard};
\node[anchor=east] at (-6.35, 2.85) {$C^\dagger \, U_{\text{MG}} \, C$};

\node[trapezium,
    draw, text centered, fill = white,  trapezium left angle=45, trapezium right angle=45,  text width = 347, text height = 9] (t) at (0, 2.08) {Provably Hard};
\node[anchor=east] at (-7.6, 2.25) {$C \, U_{\text{MG}} $};

 \node[trapezium,
    draw, text centered, fill = white,  trapezium left angle=45, trapezium right angle=45,  text width = 385.5, text height = 9] (t) at (0, 1.393) {CI Restricted Bitstring Output (CIbO)};
\node[anchor=east] at (-7.6, 1.53) {$C_{\text{P}}^\dagger \, U_{\text{MG}} \, C_{\text{P}}$};

\node[trapezium,
    draw, text centered, fill = white,  trapezium left angle=45, trapezium right angle=45,  text width = 425, text height = 9] (t) at (0, 0.699) {Computational IBO (CIBO)};
\node[anchor=east] at (-8.05, 0.8) {$U_{\text{CZ}}^\dagger \, U_{\text{MG}} \, U_{\text{CZ}}$};

\node[trapezium,
    draw, text centered, fill = white,  trapezium left angle=45, trapezium right angle=45,  text width = 465, text height = 9] (t) at (0,0) {Product Input Bitstring Output (PIBO)};
\node[anchor=east] at (-9.6, 0.1) {$S^{\dagger} U_{\text{MG}} S$};

\draw[thick,->] (10, -0.1) -- (10, 3);

\node[anchor=east, text width = 75, align = center] at (9.9, 2) {Loss of Simulability};

\end{tikzpicture}
}
% \end{center}
  \caption{A visual guide to the hierarchy of Clifford conjugated matchgate circuits we are going to consider. As we go up the hierarchy, {\color{edits} we allow for the action of Cliffords to act more freely. Inversely, as we allow for more Cliffords to act we lose versions of simulability. We begin with circuits for which we can compute all bitstring outputs and marginals on various inputs. As we go up, we lose inputs/outputs for which marginals are simulable, and eventually lose marginals altogether. We end} with circuits for which only a polynomial number of Pauli expectation values are simulable, {\color{edits} and for which we conjecture sampling is hard}. {\color{edits} $S$ represents SWAP circuits,} {\color{edits}$U_{\text{CZ}}$ } represents control-$Z$ circuits, $C_{\text{P}}$ arbitrary Clifford permutations, and $C$ arbitrary Cliffords. }
\end{figure*}

\subsection{SWAP Conjugated Matchgate Circuits}

\noindent \textbf{Theorem 1.}
\label{JWPS} {\color{edits} SWAP Conjugated Matchgates.}
    \begin{enumerate}
        \item {\color{edits} SWAP conjugated matchgates are PIBO simulable (Product-input bitstring-output). }
        \item The only ancilla-free encodings such that the generators required to make arbitrary product state are within $\mathcal{L}_{1,2}^{(n)}$ are the Jordan-Wigner encoding {\color{edits} SWAP conjugated Jordan-Wigner encoding. There is also the freedom to conjugate by a single local Clifford on one specified qubit}.  
    \end{enumerate}

{\color{edits} \noindent \textit{Proof of statement 1}: Via classical processing, this statement is trivial. The action of SWAP conjugation can be absorbed into {\color{edits} the} input product state, and desired bitstring output. SWAP gates map product states to product state, and the action of SWAP on {\color{edits} the} output just permutes which qubits we are taking output over. \\}

\noindent \textit{Proof of statement 2}: 

  \indent We will show how to make a product state with generators of an anticommuting set of Pauli strings and their quadratic products $\mathcal{L}_{1,2}^{(n)}$. We will show that this must be the Jordan-Wigner encoding or a SWAP conjugated Jordan-Wigner encoding, with minor additional freedoms.  We have freedom of the addition of the total parity operator $Z_1 Z_2 Z_3 ...Z_n$ and the freedom to conjugate {\color{edits} the} first qubit by any single qubit Clifford.  \\ \\
\indent This is a proof by construction: we will construct a set of anticommuting Pauli strings $\mathcal{L}_1^*$ (and the union of those Paulis and their quadratic products $\mathcal{L}_{1,2}^*$), and show that $\mathcal{L}_1^* = \mathcal{L}_1^{(n)}$ for the Jordan-Wigner or conjugated Jordan-Wigner encoding. We prove this Theorem in four Propositions. 

\noindent \textbf{Proposition 1.} \label{prop1}
    The generators of arbitrary single qubit rotations cannot be in $\mathcal{L}_{1,2}^*$ for more than two qubits. 
\begin{proof}
       To make a product state on qubit $1$ (without loss of generality for some other index) we will require access to at least two of the three generators $X_i$, $Y_i$, and $Z_i$. To have access to two of the three generators $X_1$, $Y_1$, and $Z_1$ in $\mathcal{L}_{1,2}^*$, there are two possibilities. Either two are in $\mathcal{L}_{1}^*$ (and the other necessarily in $\mathcal{L}_{2}^*$), or all are in $\mathcal{L}_{2}^*$ i.e. are quadratic products an encoding's Majoranas.  \\
\indent Without loss of generality, we will consider the former in our construction. We leave it as an exercise to the reader to assume the latter. We will define arbitrary local single Pauli operators on qubit one as
\begin{align*}
    \tilde X &= \text{Choose 1 from } \{X,Y,Z \} {\color{edits},} \\
    \tilde Y &= \text{Choose 1 from } \{X,Y,Z \} \symbol{92} \{\tilde X\} {\color{edits},} \\
    \tilde Z &= \tilde X \tilde Y {\color{edits},}
\end{align*}
where $\tilde X_1, \tilde Y_1 \in \mathcal{L}_{1}^*$ and $\tilde Z_1 \in \mathcal{L}_{2}^*$. We cannot add local generators of some other qubit (without loss of generality qubit 2) to $\mathcal{L}_{1}^*$ without breaking anticommutation structure, so the only way to add local generators is by adding all Pauli strings $X_2$, $Y_2$, $Z_2$ to $\mathcal{L}_{2}^*$. \\
\indent This requires Majoranas {\color{edits} that anti-coomute with $\tilde X$ and $\tilde Y$},
\begin{align*}
    c_x = \tilde Z \otimes X \otimes p_3 \otimes p_4 \otimes ... \, p_n \ \in \mathcal{L}_1^* {\color{edits},} \\
    c_y = \tilde Z \otimes Y \otimes p_3 \otimes p_4 \otimes ... \, p_n \ \in \mathcal{L}_1^* {\color{edits},} \\
    c_z = \tilde Z \otimes Z \otimes p_3 \otimes p_4 \otimes ... \, p_n \ \in \mathcal{L}_1^* {\color{edits},}
\end{align*}

where $p_k$ are some single set of arbitrary Pauli letters $\{X,Y,Z,I \}$ for $3 \leq k \leq n$. \\
\indent We now present a proof by contradiction, that we {\color{edits} also} cannot have {\color{edits} any pair of} $X_3$, $Y_3$, and $Z_3$ in $\mathcal{L}_2^*$, due to restrictions on sets of mutually commuting and anticommuting Paulis. \\
\indent We could not add the local generators to $\mathcal{L}_{1}^*$ without breaking the required anticommutation structure. Thus, let's assume $X_3$, $Y_3$, and $Z_3$ are in $\mathcal{L}_2^*$. This requires Majoranas
\begin{align*}
    c_x' &= \tilde Z \otimes p_2' \otimes X \otimes p_4' \otimes ... \ p_n' \ \in \mathcal{L}_1^* {\color{edits},} \\
    c_y' &= \tilde Z \otimes p_2' \otimes Y \otimes p_4' \otimes ... \ p_n' \ \in \mathcal{L}_1^* {\color{edits},} \\
    c_z' &= \tilde Z \otimes p_2' \otimes Z \otimes p_4' \otimes ... \ p_n' \ \in \mathcal{L}_1^* {\color{edits},}
\end{align*}
 such that $X_3$ = $c_z' c_y'$, and likewise for the other single qubit Paulis. We note $p_k'$ are some other set of Pauli letters again for $2 \leq k \leq n$ ($k \not = 3$). We require that these six Pauli strings $\{c_x, c_y, c_z, c_x', c_y', c_z'\}$ anticommute. Since all six Pauli strings have the same letter $\tilde Z$ on the first qubit, this means the six truncated Pauli strings on qubits $2$ through $n$ must anticommute.\\
% \begin{align*}
    % &\{\tilde Z \, X \, p_3 \, p_4..., \tilde Z \, Y \, p_3 \, p_4... , \tilde Z \, Z \, p_3 \, p_4..., \\
    % &\tilde Z \, p_2' \, X \, p_4' ..., \tilde Z \, p_2' \, Y \, p_4'... , \tilde Z \, p_2' \, Z \, p_4'...\}
%     \{c_x, c_y, c_z, c_x', c_y', c_z'\}
% \end{align*} 
% all mutually anticommute. 
\indent Consider the two truncated Pauli strings made up of the Pauli letters $\{ p_k \}$ and $\{p_k' \}$ on qubits 4 onward: 
\begin{equation}
q \equiv p_4 \otimes p_5 \otimes ... \ p_n  {\color{edits},} \qquad q' \equiv p_4' \otimes p_5' \otimes ... \ p_n' {\color{edits}.}
\end{equation}
These length $n-3$ Pauli strings $q, q'$ either commute or anticommute. If they commute (anticommute), then it must be that the remainder Pauli strings given by taking the letters of the six Majoranas over qubits 2 and 3, 
\begin{align*}
R \, \equiv \, \{ \ &X \otimes p_3 , \ Y \otimes p_3 , \ Z \otimes p_3, \\
&p_2' \otimes X, \ p_2' \otimes Y, \ p_2' \otimes Z \ \ \}
\end{align*} 
mutually anticommute (commute). This, however, is impossible: over two qubits, the maximum number of mutually anticommuting (commuting) Pauli strings is five (four)\cite{anticomm}. Thus, if  $X_1$, $Y_1$, and $Z_1$ are in $\mathcal{L}_2^*$, we cannot simultaneously have $X_3$, $Y_3$, and $Z_3$ in $\mathcal{L}_{2}^*$. \\
\indent Thus, the generators of arbitrary single qubit rotations cannot be in $\mathcal{L}_{1,2}^*$ for more than two qubits. 
\end{proof}
\noindent \textbf{Proposition 2.}
\label{prop2}
    % Assume prior construction from Proposition \ref{prop1a}. The swapping operations required to make product states on qubits without access to arbitrary single qubit rotations implies $XX$ and $YY$ are in $\mathcal{L}_{1,2}^*$. 
    {\color{edits} Access to some swap-like gate implies} some $XX$ and $YY$ generators are in $\mathcal{L}_{1,2}^*$. 
\begin{proof}
     In our proof by construction so far, we have
\begin{align*}
    \tilde X_1, \  \tilde Y_1 \ &\in \ \mathcal{L}_1^{*} {\color{edits},} \\
    \tilde Z \otimes X \otimes p_3 \otimes p_4 \otimes ... \ p_n \ &\in \ \mathcal{L}_1^{*}  {\color{edits},}\\
    \tilde Z \otimes Y \otimes p_3 \otimes p_4 \otimes ... \ p_n \ &\in \ \mathcal{L}_1^{*}  {\color{edits},} \\
    \tilde Z \otimes Z \otimes p_3 \otimes p_4 \otimes ... \ p_n \ &\in \ \mathcal{L}_1^{*} {\color{edits}.} 
\end{align*} \\
\indent Without the ability to have local rotations on any other qubit, the only way to make arbitrary product states on the other qubit states is to first make them on states {\color{edits} the} first qubits, and then swap that state to the qubits for which we have no access to local rotations. Our swapping operation will have two conditions: 
\begin{align}
    U_{\text{SWAP}} (a|0\rangle + b|1\rangle)|0\rangle &= |0\rangle(a|0\rangle + b|1\rangle) \\
    U_{\text{SWAP}}|0\rangle(a|0\rangle + b|1\rangle) &=(a|0\rangle + b|1\rangle)|0\rangle {\color{edits}.} 
\end{align}
We will assume that any gate which performs this function will do so regardless of amplitudes on the product state. There are two gates which satisfy these two conditions. We can solve and find
\begin{equation}
    U_{\text{SWAP}} = \begin{pmatrix}
        1 & 0 & 0 & 0 \\
        0 & 0 & 1 & 0 \\
        0 & 1 & 0 & 0 \\
        0 & 0 & 0 & \pm 1
    \end{pmatrix}
\end{equation}
are the only two gates that satisfy this condition; we will label these two gates as SWAP and fSWAP accordingly. We recall that
\begin{align}
    \text{SWAP} &= I + XX + YY + ZZ = ie^{-\frac{i \pi}{4}(I + XX + YY + ZZ)}{\color{edits},} \\
    \text{fSWAP} &= XX + YY + IZ + ZI = i e^{-\frac{i \pi}{4} ( XX + YY + IZ + ZI)}{\color{edits}.}
\end{align}
As our set needs to be closed under commutation, we have that the generators showing up in the exponential sum of Pauli terms implies those generators are in $\mathcal{L}_{1,2}^{*}$. For either gate we could have access to, we require $XX$ and $YY$ in $\mathcal{L}_{1,2}^{*}$. \\
\end{proof}

\noindent \textbf{Proposition 3.} \label{prop3}
    % Assume prior construction from Proposition \ref{prop1}. We cannot have access to swapping operations from both qubits with access to arbitrary single qubit rotations to the same qubit, unless our system size is $n=3$. 
    Given the results of Proposition \hyperref[prop1]{1} and Proposition  \hyperref[prop1]{2}, there are limits to what $XX$ and $YY$ can be simultaneously present in $\mathcal{L}_{1,2}^*$.
    We cannot have access to swap-like operations from both qubits with access to arbitrary single qubit rotations to the same qubit, unless {\color{edits} the} system size is $n=3$. 
\begin{proof}
    We will show two cases: if we have access to swapping operations from qubit $1$ to $3$, we cannot also have swapping operations from $2$ to $3$ unless $n=3$. Similarly, if we have access to swapping operations from qubit $2$ to $3$, we cannot also have swapping operations from $1$ to $3$ unless $n=3$. {\color{edits} Notation and definition of terms follows from Propositions \hyperref[prop1]{1} and \hyperref[prop2]{2}}. \\
    \indent \textit{Assume we desire swapping operations from qubit $1$ to $3$. } Since $X_1 X_3$ and $Y_1 Y_3$ commute, they cannot both be in $\mathcal{L}_{1}^{*}$ as all operators anticommute. Either one is in $\mathcal{L}_{1}^{*}$, or both are in $\mathcal{L}_{2}^{*}$.
    \begin{itemize}
        \item Assuming that $X_1 X_3 \in \mathcal{L}_{1}^{*}$ (without loss of generality for  $Y_1 Y_3 \in \mathcal{L}_{1}^{*}$). To mutually anticommute with $\tilde X, \tilde Y$, it must be that $\tilde X, \tilde Y = Y_1, Z_1$ accordingly. For $Y_1 Y_3$ to be in $\mathcal{L}_{2}^{*}$, knowing that $Y_1 Y_3$ and $Z_1$ anticommute and $Z_1 \in \mathcal{L}_1^{*}$, it must be that $Y_1 Y_3 = Z_1 \cdot c$ for $c$ a Majorana which anticommutes with $Z_1$. The only option is that this $c = X_1 Y_3$. We thus have added two Majoranas to our set, $X_1 Y_3$ and $X_1 X_3$, which also fixes $p_3 = Z$.
        \item  For $X_1 X_3$ and $Y_1 Y_3$ both in $\mathcal{L}_{2}^{*}$, regardless of choice of Paulis for $\tilde X, \tilde Y$ it must be that $\tilde Y$ anticommutes with $X_1 X_3$ and $\tilde X$ anticommutes with $Y_1 Y_3$. We leave it as an exercise to the reader to show that this implies $\{ \tilde X, \tilde Y \} = \{ X, Y \}$ and that we add Majoranas $Z_1 Y_3$ and $Z_1 X_3$ to our set. In any scenario, $p_3 = Z$. 
    \end{itemize} 
\indent {\color{edits} Either way, $p_3 = Z$. Thus,} without loss of generality assume that $X_1 X_3$ and $Y_1 Y_3$ are both in $\mathcal{L}_{2}^{*}$ {\color{edits} and }$\{ X_1, Y_1\} \in \mathcal{L}_{1}^{*}$. We then have
\begin{align*}
    X_1, \ Y_1 & \ \in \ \mathcal{L}_{1}^*{\color{edits},} \\
    Z \otimes X \otimes Z \otimes p_4 \otimes ... \ p_n \ & \ \in  \ \mathcal{L}_{1}^* {\color{edits},}\\
    Z \otimes Y \otimes Z \otimes p_4 \otimes ... \ p_n \ & \ \in \ \mathcal{L}_{1}^* {\color{edits},}\\
    Z \otimes Z \otimes Z  \otimes p_4 \otimes ... \ p_n & \ \in \ \mathcal{L}_{1}^* {\color{edits},}\\
    Z_1 X_3, \ Z_1 Y_3 & \ \in \  \mathcal{L}_{1}^*{\color{edits}.}
\end{align*} If we attempt to add $X_2 X_3$ and $Y_2 Y_3$, we immediately have that both must be in $\mathcal{L}_{2}^{*}$. Since $Z \otimes X \otimes Z \otimes p_4 \otimes ... \ p_n$ must anticommute with $X_2 X_3$ and $Z_1 Y_3$, it must be that 
\begin{align*}
    X_2 X_3 = (Z \otimes X \otimes Z \otimes p_4 \otimes ... p_n) \cdot (Z_1 Y_3)
\end{align*}
which is only possible if $ p_4,p_5,...,p_n $ are all the identity. Thus, to have both sets in $\mathcal{L}_{1,2}^{*}$ restricts us to only having three qubits that we act on non-trivially, i.e with dynamics not equivalent to the identity. \\
\indent\textit{\textit{Assume we desire swapping operations from qubit $2$ to $3$. } } 
    With $\tilde X$ and $\tilde Y$ in $\mathcal{L}_{1}^{*}$ it is obvious that $ X_2 X_3, Y_2 Y_3 \not \in \mathcal{L}_{1}^{*}$ due to anticommutation. Thus, these two terms can only be in $\mathcal{L}_{2}^{*}$. We will start by only considering $X_2 X_3$. To be in $\mathcal{L}_{2}^{*}$, there must exist elements in $\mathcal{L}_{1}^{*}$
\begin{align*}
    c_{p_1} &= \tilde Z \otimes p_2' \otimes p_3' \otimes p_4^* \otimes p_5^* \otimes... \ p_n {\color{edits},}\\
    c_{p_2} &= \tilde Z \otimes p_2'' \otimes p_3'' \otimes p_4^* \otimes p_5^* \otimes... \ p_n{\color{edits}.} \\
\end{align*}
that multiply together to $X_2 X_3$ while anticommuting. This leaves only a finite number of options for what the Pauli strings $p_2' p_3'$ and $p_2'' p_3''$ can be, with the four options 
\begin{align*}
    \{ \ p_2'\otimes p_3', \ p_2'' \otimes p_3'' \} \ := \ \ & \{ ZI, YX \}, \{IZ, YX\},\\
    & \{YI,ZX\},\{ IY, XZ \}{\color{edits}.}
\end{align*}
To have $c_{p_1}, c_{p_2}$ in $\mathcal{L}_{1}^{*}$, it must be that 
\begin{align*}
    \{ \ \tilde Z\otimes p_2'\otimes p_3' \otimes p_4^* \otimes p_5^* \otimes ... \ &p_n, \\
    \tilde Z \otimes p_2''\otimes p_3''\otimes p_4^* \otimes p_5^* \otimes ... \ &p_n, \\
    \tilde Z \otimes X \otimes p_2 \otimes p_3 \otimes p_4 \otimes ... \ &p_n,\\
 \tilde Z \otimes  Y \otimes p_2 \otimes p_3 \otimes p_4 \otimes ... \ &p_n,\\
\tilde Z \otimes Z \otimes p_2 \otimes p_3 \otimes p_4 \otimes ... \ &p_n  \ \ \}
\end{align*}
all mutually anticommute. If we assume $p_3 \otimes p_4 \otimes ...p_n$ and $ p_4^* \otimes p_5^* \otimes ...p_n$ anticommute, then we have a contradiction regarding the size of mutually commuting sets \cite{anticomm}. Assuming $p_3 \otimes p_4 \otimes ...p_n$ and $ p_4^* \otimes p_5^* \otimes ...p_n$ commute however, we require the truncated Pauli strings $\{p_2'\otimes p_3', p_2'' \otimes p_3'',X \otimes p_2, Y \otimes p_2, Z \otimes p_2  \}$ to mutually anticommute. \\
\indent While there is no contradiction due to the size of mutually anticommuting sets of Pauli strings, for all possible choices of $p_2' \otimes p_3', p_2'' \otimes p_3''$ there is no way to construct a set of mutually anticommuting Pauli strings. \\
\indent The only way to resolve this is to choose 
\begin{align*}
    p_2'\otimes p_3' = IY{\color{edits},} \qquad  p_2'' \otimes p_3'' =  XZ{\color{edits},}
\end{align*}
and have that $$\tilde Z\otimes X \otimes Z \otimes p_4^*\otimes p_5^* ...p_n = \tilde Z\otimes X\otimes p_3 \otimes p_4 \otimes p_5 \otimes ...p_n,$$ thus fixing $p_3 = Z$. A similar proof follows when considering $Y_2 Y_3$. This leaves us with 
\begin{equation*}
\begin{aligned}
    \tilde X_1, \tilde Y_1 & \in  \mathcal{L}_{1}^{*} {\color{edits},}\\
     \tilde Z \otimes X \otimes Z \otimes p_4 \otimes p_5 \otimes ... p_n, & \in  \mathcal{L}_{1}^{*} {\color{edits},}\\
         \tilde Z \otimes Y \otimes Z \otimes p_4 \otimes p_5 \otimes ... p_n, & \in  \mathcal{L}_{1}^{*}{\color{edits},}\\
              \tilde Z \otimes Z \otimes Z \otimes p_4 \otimes p_5 \otimes ... p_n, & \in  \mathcal{L}_{1}^{*} {\color{edits},} \\
     \tilde Z \otimes I \otimes X \otimes p_4 \otimes p_5 \otimes ... p_n, & \in  \mathcal{L}_{1}^{*} {\color{edits},} \\
     \tilde Z \otimes I \otimes Y \otimes p_4 \otimes p_5 \otimes ... p_n, & \in  \mathcal{L}_{1}^{*}{\color{edits}.}
\end{aligned}
\end{equation*}
We can now consider what happens if we attempt to add $X_1 X_3$ and $Y_1 Y_3$ to $\mathcal{L}_{1,2}^{*}$. With any choice of $\tilde X$ and $\tilde Y$, $X_1 X_3$ and $Y_1 Y_3$ must be in $\mathcal{L}_{2}^{*}$. For $ \{\tilde X, \tilde Y \} = \{ Y, Z \}$ or $\{X,Z \}$, there is no way to add Majoranas that allow $X_1 X_3$ and $Y_1 Y_3$ to be in $\mathcal{L}_{2}^{*}$. For $\{ \tilde X, \tilde Y \} = \{ X, Y \}$, anticommutation fixes that
\begin{align*}
    X_1 X_3 &= Y_1 \cdot (Z \otimes I \otimes X \otimes p_4 \otimes p_5 \otimes ... \ p_n ) {\color{edits},}\\
    Y_1 Y_3 &= X_1 \cdot (Z \otimes I \otimes Y \otimes p_4 \otimes p_5 \otimes ... \ p_n){\color{edits},}\\
\end{align*}
which can only happen if $p_4, p_5, ...,p_n$ are all the identity, and we have no other qubits in our system. \\  
Thus, in either scenario, having swapping operations from both qubits with access to arbitrary single qubit rotations to the same qubit fixes system size $n=3$. 
\end{proof}

\noindent \textbf{Proposition 4.}
    % Assume prior construction from Proposition \ref{prop3a}. To add swapping operations from qubit $n$ to $n+1$ fixes that Majoranas in $\mathcal{L}_1^*$ are products of $Z$ strings, with Pauli $X$ and $Y$ only one one index. 
    Given the result of Proposition \hyperref[prop3]{3}, there are only certain set{\color{edits}s} of Majoranas such that the necessary $XX$ and $YY$ generators are in $\mathcal{L}_{1,2}^*$.
    The ability to do swap-like operations from qubit $n$ to $n+1$ fixes that Majoranas in $\mathcal{L}_1^*$ are products of $Z$ strings, with Pauli $X$ and $Y$ only one one index.

\begin{proof}
    {\color{edits} Notation and definition of terms follows from Propositions \hyperref[prop1]{1}, \hyperref[prop2]{2}, and \hyperref[prop3]{3}}. Without loss of generality for possible constructions, assume that $X_1 X_3$ and $Y_1 Y_3$ are both in $\mathcal{L}_{2}^{*}$, and
\begin{align*}
    X_1, \ Y_1 \ & \ \in  \mathcal{L}_{1}^{*}{\color{edits},} \\
    Z \otimes X \otimes Z \otimes p_4 \otimes ... \ p_n \ & \ \in  \mathcal{L}_{1}^{*}{\color{edits},} \\
    Z \otimes Y \otimes Z \otimes p_4 \otimes ... \ p_n \ & \ \in  \mathcal{L}_{1}^{*}{\color{edits},} \\
    Z \otimes Z \otimes Z \otimes p_4 \otimes ... \ p_n \ & \ \in   \mathcal{L}_{1}^{*} {\color{edits},}\\[0.2em]
    Z_1 X_3, \ Z_1 Y_3 \ & \ \in  \mathcal{L}_{1}^{*}{\color{edits}.}
\end{align*}

Rather than building swapping operations from qubits $3$ to $4$, we will instead construct the remaining terms via induction. Assume that we have defined operators over the first $n$ qubits, and that the following terms are in $\mathcal{L}_1^{(n+1)}$. 
\begin{align*}
    Z \otimes X \otimes Z \otimes Z \otimes ... \ Z & \otimes p_{n+1} \  \in  \mathcal{L}_{1}^{(n+1)}{\color{edits},} \\
     Z \otimes Y \otimes Z \otimes Z \otimes ... \ Z & \otimes p_{n+1} \  \in  \mathcal{L}_{1}^{(n+1)}{\color{edits},} \\
     Z \otimes Z \otimes Z \otimes Z \otimes ... \ Z & \otimes p_{n+1} \ \in  \mathcal{L}_{1}^{(n+1)}{\color{edits},} \\
          Z \otimes I \otimes Z \otimes Z \otimes ... \ X & \otimes I \ \quad \, \, \in  \mathcal{L}_{1}^{(n+1)}{\color{edits},} \\
    Z \otimes I \otimes Z \otimes Z \otimes ... \ Y & \otimes I \ \quad \, \, \in  \mathcal{L}_{1}^{(n+1)}{\color{edits}.} \\
    %  Z \otimes I \otimes Z \otimes Z \otimes ... \ X & \ \in  \mathcal{L}_{1}^{(n+1)} \\
    % Z \otimes I \otimes Z \otimes Z \otimes ... \ Y & \ \in  \mathcal{L}_{1}^{(n+1)} \\
\end{align*}
We note that for $n=3$, we already have the base case. We now ask about the conditions for $X_n X_{n+1}$ and $Y_n Y_{n+1}$ to be in $\mathcal{L}_2^{(n+1)}$. Just considering $X_n X_{n+1}$ (the proof for $Y_n Y_{n+1}$ follows near identically), anticommutation of $X_n X_{n+1}$ and 
\begin{align*}
    Z \otimes I \otimes Z \otimes Z \otimes ... \ Y \ \in  \mathcal{L}_{1}^{(n+1)} 
\end{align*}
implies $X_n X_{n+1}$ is a product which includes that Majorana. We thus can write
\begin{align*}
    X_n X_{n+1} \ = \ (& Z \, \otimes \, I \, \otimes \, Z \, \otimes \, Z \, \otimes ... \, \otimes \,  Y \ )\\ 
    \cdot \ (& p_1 ' \otimes p_2' \otimes p_3' \otimes p_4' \otimes ... \, \otimes \,  p_{n+1}')
\end{align*}
which fixes the additional undetermined Majorana to be 
\begin{equation}
    Z \otimes I \otimes Z \otimes Z \otimes ... \ Z \otimes X \in \mathcal{L}_1^{(n+1)}{\color{edits}.}
\end{equation}
Thus, via induction, we see {\color{edits} the} Majoranas are necessarily products of Pauli $Z$ terms with growing locality, with the final Pauli letter in each string being $X$ or $Y$. 
\end{proof}
    By combining Propositions 1 through 4, we conclude our proof. Choosing any possible construction of swapping operations to get product states on $n$ qubits, we find that {\color{edits} the} final constructed set $\mathcal{L}_{1}^{*}$ looks something like 
\begin{equation}
\begin{aligned}
    \mathcal{L}_{1} &= \{\tilde X, \ \tilde Y, \ \tilde Z \otimes XZZZ..., \ \tilde Z \otimes YZZZ..., \\
    & \tilde Z \otimes ZZZZ..., \ \tilde Z \otimes IX..., \ \tilde Z \otimes IY...,... \}{\color{edits}.}
\end{aligned}
\end{equation}
Any possible way of building {\color{edits} a set which can generate product states} looks like Jordan-Wigner or SWAP conjugated Jordan-Wigner, with the additional freedom that any local Clifford can act on the qubit for which two of the three local Pauli generators are in $\mathcal{L}_{1}^{(n)}$. 

\subsection{CZ+SWAP Conjugated Matchgates} 

\noindent \textbf{Theorem 2.}
\label{thm:CZM}
    {\color{edits} CZ+SWAP conjugated matchgate circuits}
    \begin{enumerate}
        \item  {\color{edits} CZ+SWAP conjugated matchgate circuits are CIBO (Computational input, bitstring output) simulable}
        \item If $Z_i\in \mathcal{L}_{2}^{(n)}$ for all $i$, then the ancilla-free encoding is connected to Jordan-Wigner by a circuit composed of only SWAP and CZ gates (we will denote the family of Cliffords generated by SWAPs and $CZ$ gates be defined as $C_{\text{CZ}}$).
        \item Simulability of bitstring outputs of $CZ$ conjugated matchgate circuits on product state inputs {\color{edits} implies} the ability to simulate bitstring outputs of arbitrary quantum circuits {\color{edits} and a collapse of the polynomial hierarchy to its third level. }
    \end{enumerate}

\noindent \textit{Proof of Statement 1.} 
We will consider the simulability task of some input computational bitstring $|x\rangle$ of Hamming weight $m$ i.e. $a_{i_m}^{\dagger}...\, a_{i_1}^{\dagger}|\mathbf{0}\rangle$, some $k$-bit output bitstring $y^*$, and unitary gate $U_{\text{G}} = {\color{edits}C_{\text{CZ}}}^{\dagger} U_{\text{MG}} {\color{edits}C_{\text{CZ}}}$. \\
\indent For writing {\color{edits} the} input state, we can observe that for any encoding defined by some Clifford in $C_{Z2}$, we can obtain raising and lowering operators $c_i^X \pm i c_i^Y \mapsto a_i,a_i^{\dagger}$. These operators look similar to the Jordan-Wigner creation and annihilation operators, except for having the products of Pauli $Z$ {\color{edits} factors appear} in different locations. We can write 
\begin{align*}
    a_k, a_k^{\dagger} \mapsto \left(\prod_{j=1}^{k-1}z_j\right) (X_k \pm iY_k) \left(\prod_{j=k+1}^{n}z_j\right) {\color{edits},}
\end{align*}
where each $z \in \{I, Z \}$. Thus, for any computational basis state input $|x\rangle$, the input can be written the exact same way with respect to the new encodings as it would have been in Jordan-Wigner, up to a sign difference. Said in another way, $CZ|x\rangle = \pm|x\rangle$. \\

\indent The $k$-bit projector $|y^*\rangle \langle y^*|$ can be written using the Jordan-Wigner Majoranas as a product of number operators {\color{edits} $a_i^{\dagger}a_i$} and {\color{edits}$1-a_i^{\dagger}a_i$}, depending on the value of $y^*_i \in \{0, 1\}$. We note that conjugation by $CZ$ circuits preserves the number operator, and thus 
\begin{equation*}
    U_{\text{CZ}}|y^*\rangle \langle y^*| U_{\text{CZ}}^{\dagger} = |y^*\rangle \langle y^*|.
\end{equation*} The problem of finding a many qubit output of a $CZ$ conjugated matchgate circuit can be mapped to
\begin{align}
    p(y^*) &= \langle x |U_{\text{CZ}}^\dagger U_{\text{MG}} \, U_{\text{CZ}} |y^*\rangle \langle y^* | U_{\text{CZ}}^\dagger U_{\text{MG}} U_{\text{CZ}}| x \rangle \\
    &= \langle \mathbf{0}|a_{i_1}... \, a_{i_m} U_{\text{MG}}^{\dagger} |y^*\rangle \langle y^*| U_{\text{MG}} \,  a_{i_m}^{\dagger}...\, a_{i_1}^{\dagger}|\mathbf{0}\rangle {\color{edits},}
\end{align}
which is a matchgate circuit which can be evaluated in polynomial time. \\ 

\noindent \textit{Proof of Statement 2.}
        We know that if $Z_i \in \mathcal{L}_{2}^{*}$ for $i \in [1,n]$, then
        \begin{align*}
            & p_1 \otimes ... \otimes p_{i-1} \otimes X \otimes p_{i+1} \otimes ... \otimes p_n \in \mathcal{L}_{1}^*{\color{edits},}\\
            & p_1 \otimes ... \otimes p_{i-1} \otimes Y \otimes p_{i+1} \otimes ... \otimes p_n \in \mathcal{L}_1^*{\color{edits}.}\\
        \end{align*}

        We can now ask about what must be true of these Pauli letters across different pairs, given that such pairs exist in $\mathcal{L}_1^*$ for all qubits.
        To do so, we will list {\color{edits} the} Majoranas in a convenient matrix

\renewcommand{\arraystretch}{1.5} % Default value: 1

\begin{equation}
\begin{pmatrix}
    c_1^{X,Y} \\ c_2^{X,Y} \\ c_3^{X,Y} \\ c_4^{X,Y} \\ ... 
\end{pmatrix}
= \begin{pmatrix}
    \{X,Y\} & p_2^1 & p_3^1 & p_4^1 & ... \\
    p_1^2 & \{X,Y\} & p_3^2 & p_4^2 & ... \\
    p_1^3 & p_2^3 & \{X,Y\} & p_4^3 & ... \\
    p_1^4 & p_2^4 & p_3^4 & \{X,Y\} & ... \\
    \vdots & \vdots & \vdots & \vdots & \ddots \\
\end{pmatrix}{\color{edits}.}   
\end{equation}
To prove $p_i^j$ cannot be $X$ nor $Y$, let us consider $p_2^1$ to be $X$ (without loss of generality $Y$), to arrive at a contradiction regarding the possible sets of mutually anticommuting Paulis. Let
\begin{equation}
\begin{pmatrix}
    c_1^{X,Y} \\ c_2^{X,Y} \\ ... 
\end{pmatrix}
= \begin{pmatrix}
    \{X,Y\} & X & p_3^1 & p_4^1 & ... \\
    p_1^2 & \{X,Y\} & p_3^2 & p_4^2 & ... \\
    \vdots & \vdots & \vdots & \vdots & \ddots \\
\end{pmatrix}{\color{edits}.} 
\end{equation}
We will now consider the (anti)commutation of $c_1^{X,Y}$ with $c_2^X$ and $c_2^Y$ separately. 
\begin{itemize}
    \item For the commutation of $c_1^{X,Y}$ with $c_2^X$, knowing that the two commute at index $2$, they must overall anticommute across all other indices. Thus, it must be that the Pauli terms $ \{X,Y \} \otimes  p_3^1 \otimes p_4^1 \otimes ... p_n^1 \text{ and } p_1^2 \otimes p_3^2 \otimes p_4^2 \otimes ... p_n^2$ anticommute. 
    \item For the commutation of $c_1^{X,Y}$ with $c_2^Y$, we know the two anticommute at index $2$. Thus, on the rest of indices, they must overall commute. Thus, $ \{X,Y \} \otimes  p_3^1 \otimes p_4^1 \otimes ... p_n^1 \text{ and } p_1^2 \otimes p_3^2 \otimes p_4^2 \otimes ... p_n^2$  must all commute.
\end{itemize}
  However, we have reached a contradiction. The terms considered above cannot anticommute and commute at the same time. Thus, letting $p_2^1$ be $X$ or $Y$ is impossible. \\
\indent This contradiction does not exist if $p_2^1 = I$ or $Z$. By considering each pairing of $c_i^{X,Y}$ it is possible to prove that \textit{no} $p_i^j$ can be anything except for $I$ or $Z$. \\
\indent Now the only restriction is assigning $Z$ and $I$ such that all terms mutually anticommute. We can further construct our encoding by only considering $4$ Majoranas at a time
\begin{equation}
\begin{pmatrix}
    c_1^{X,Y} \\ c_2^{X,Y} \\ ... 
\end{pmatrix}
= \begin{pmatrix}
    \{X,Y\} & p_2^1 & p_3^1 & p_4^1 & ... \\
    p_1^2 & \{X,Y\} & p_3^2 & p_4^2 & ... \\
    \vdots & \vdots & \vdots & \vdots & \ddots \\
\end{pmatrix}{\color{edits}.} 
\end{equation}
Knowing that $p_i^j$ are either $Z$ or $I$, we know the only places that $c_1^{X,Y}$ and $ c_2^{X,Y} $ anticommute are in indices $1$ and $2$, and thus we only need to consider the terms $p_2^1$ and $p_1^2$. If both terms are either $Z$ or $I$, $c_1^{X,Y}$ and $ c_2^{X,Y} $ commute. Thus, if $p_1^2$ is $Z$, that $p_2^1$ is $I$ (and vice versa). \\
\indent By considering all pairs of $c_i$, if we set $p_i^j = Z$, $p_j^i = I$. \\
 
\indent We can now consider all encodings for which $Z_j \in \mathcal{L}_{2}^{n}$ for all $j$, by observing all possible matrices which encode the valid operators. This includes Jordan-Wigner {\color{edits} with the} choice 
\begin{equation}\label{JWMat}
\begin{pmatrix}
    c_1^{X,Y} \\ c_2^{X,Y} \\ c_3^{X,Y} \\ c_4^{X,Y} \\ ... 
\end{pmatrix}
= \begin{pmatrix}
    \{X,Y\} & I & I & I & ... \\
   Z & \{X,Y\} & I & I & ... \\
    Z & Z & \{X,Y\} & I & ... \\
    Z & Z & Z & \{X,Y\} & ... \\
    \vdots & \vdots & \vdots & \vdots & \ddots \\
\end{pmatrix}   {\color{edits}.} 
\end{equation}
We can see a choice like 
\begin{equation}
\begin{pmatrix}
    c_1^{X,Y} \\ c_2^{X,Y} \\ c_3^{X,Y} \\ c_4^{X,Y} \\ ... 
\end{pmatrix}
= \begin{pmatrix}
    \{X,Y\} & Z & Z & Z & ... \\
   I & \{X,Y\} & I & I & ... \\
    I & Z & \{X,Y\} & Z & ... \\
    I & Z & I & \{X,Y\} & ... \\
    \vdots & \vdots & \vdots & \vdots & \ddots \\
\end{pmatrix}  {\color{edits}.}  
\end{equation}
is SWAP and re-ordering equivalent to Jordan-Wigner. By swapping qubits $\{1,4\}$, then $\{1,2 \}$, it only requires fSWAPs (or just relabeling) to map back to the same matrix as Jordan-Wigner. \\
\indent However, we can also find valid encodings which cannot be mapped back to Jordan-Wigner purely with SWAP and fSWAP gates. We can, for example, consider 
\begin{equation}\label{matExam}
\begin{pmatrix}
    c_1^{X,Y} \\ c_2^{X,Y} \\ c_3^{X,Y} \\ c_4^{X,Y} \\ ... 
\end{pmatrix}
= \begin{pmatrix}
    \{X,Y\} & Z & I & I & ... \\
   I & \{X,Y\} & Z & I & ... \\
    Z & I & \{X,Y\} & Z & ... \\
    Z & Z & I & \{X,Y\} & ... \\
    \vdots & \vdots & \vdots & \vdots & \ddots \\
\end{pmatrix}   {\color{edits}.} 
\end{equation}
considering the action of all possible SWAP and fSWAP gates, there is no re-ordering of rows and columns which can map this set of spins back to Jordan-Wigner. \\
\indent For some encoding that is not SWAP/fSWAP equivalent to Jordan-Wigner, it require{\color{edits}s} conjugating {\color{edits} the} encoding by $CZ$. Recall the action of $CZ$ conjugation on Pauli strings \cite{nielsen00}, 
\begin{align*}
    CZ(X_1)CZ = XZ{\color{edits} ,}\; \; \; & CZ(Y_1)CZ = YZ{\color{edits} ,} \; \; \; CZ(X_2)CZ = ZX{\color{edits},}  \\
    CZ(Y_2)CZ = ZY{\color{edits} ,} \; \; \; & CZ(Z_2)CZ = IZ{\color{edits} ,} \; \; \; CZ(Z_1)CZ = ZI{\color{edits}.} 
\end{align*}
Thus we can see that $CZ$ effectively swaps the locations of which indices $p_{i}^j$ have label $I$ or $Z$. Taking the encoding from {\color{edits}equation} \ref{matExam} as an example. Conjugating the encoding matrix for Jordan-Wigner (\ref{JWMat}) by 
\begin{align*}
    {\color{edits} C_{\text{CZ}}} = CZ_{12} \ CZ_{23} \ CZ_{34} \ ... \ CZ_{n-1n}{\color{edits},}
\end{align*}
we derive {\color{edits}the }encoding {\color{edits}in equation} \ref{matExam}. \\
\indent While each specification defines a unique encoding, we can also derive each encoding by starting off with one matrix, and then observing the action of all possible SWAPs, fSWAPs, and $CZ$ conjugations. The action of conjugation by each can be described accordingly. 
\begin{itemize}
    \item The action of SWAP conjugation on our representation is equivalent to swapping columns.
    \item The action of fSWAP conjugation/fermionic reordering on our representation is equivalent to swapping rows.
    \item The action of $CZ$ conjugation on our representation flips whether $p_i^j = Z$ or $I$ between two rows of Majoranas. 
\end{itemize}
\textit{A brief note on perturbative simulation}

For encodings which are close to Jordan-Wigner, we have that circuits may be perturbitavely PIBO, {\color{edits} i.e PIBO with a limited number of interacting fermionic gates}. Take for example an encoding in which we fill out {\color{edits} the} matrix to be re-ordered Jordan-Wigner, except for two flips of $Z$ and $I$, 
\begingroup                       % Limits scope of arraystretch %%%%%%%%%%%%%%%%%%
\renewcommand{\arraystretch}{1.5} % Default value: 1
\begin{equation}
\begin{pmatrix}
    c_1^{X,Y} \\ c_2^{X,Y} \\ c_3^{X,Y} \\ c_4^{X,Y} \\ \vdots
\end{pmatrix}
= \begin{pmatrix}
    \{X,Y\} & I & Z & Z & ... \\
   Z & \{X,Y\} & I & Z & ... \\
    I & Z & \{X,Y\} & Z & ... \\
    I & I & I & \{X,Y\} & ... \\
    \vdots & \vdots & \vdots & \vdots & \ddots \\
\end{pmatrix}   {\color{edits}.} 
\end{equation}
For $X_iX_{i+1}$ (and $Y_iY_{i+1}$ without loss of generality),  for $i \in [4,n]$ there is no change in the fermionic resources required to build fSWAP operations. However, for $i \in \{1,2,3\}$, we can see that 
\begin{align*}
    X_1 X_2 = c_1^Y c_2^X c_3^X c_3^Y \; \; \; & X_2 X_3 = c_1^X c_1^Y c_2^Y c_3^X {\color{edits},} \\
    X_3 X_4 = c_2^X c_2^Y c_3^X c_4^Y  \; \; \;  & X_1 X_3 = c_1^X c_2^X c_2^Y c_3^Y {\color{edits},} \\
    X_1 X_4 = c_1^X c_3^X c_3^Y c_4^Y \; \; \; & X_2 X_4 = c_1^X c_1^Y c_2^X c_4^Y {\color{edits}.} 
\end{align*}
{\color{edits} In the best case,} building an arbitrary product state require{\color{edits}s} a constant scaling number of interacting fermionic gates. We can see this by building arbitrary rotations on qubit $n$, then seeing {\color{edits}that} swapping that product state down the chain from qubit $n$ to qubit $1$, then repeating this procedure for qubits $2$ and $3$ requires $6$ interacting fermionic gates. Thus, for at least some of these non Jordan-Wigner encodings, we can do perturbitavely simulable circuits on product state inputs \cite{Dias_2024, mocherla2024extending}. \\
\indent Take however another example, in which we construct our encoding matrix as follows: for each row $i$, let $p_{i+1+2k}^i = Z$, and  $p_{i+2+2k}^i = I$ for $k \in [0, \lfloor \frac{n-i}{2} \rfloor]$. {\color{edits} The encoding matrix is}
\begin{equation}
\begin{pmatrix}
    c_1^{X,Y} \\ c_2^{X,Y} \\ c_3^{X,Y} \\ c_4^{X,Y} \\ ... 
\end{pmatrix}
= \begin{pmatrix}
    \{X,Y\} & Z & I & Z & I &  ... \\
   I & \{X,Y\} & Z & I & Z & ... \\
    Z & I & \{X,Y\} & Z & I & ... \\
    I & Z & I & \{X,Y\} & Z & ... \\
    \vdots & \vdots & \vdots & \vdots & \ddots \\
\end{pmatrix} \, .  
\end{equation}
\endgroup                            %%%%%%%%%%%%%%%%%%%%%%%%%%%%%%%%%%%%
Due to the alternating pattern of $I$ and $Z$ created, generating a set of single qubit local universal generators and any set of fSWAP operations requires a linearly scaling number of interacting fermionic gates. For single qubit unitaries (without loss of generality we will study $X_i$ and $Z_i$), while $Z_i$ is made of quadratic products of Majoranas, $X_i$ requires $O(n)$, due to a need to cancel out the alternating $Z$ Pauli strings. For any $X_i X_j$, we leave it as an exercise to the reader to show that the lowest weight product of Majoranas needed is weight four, with $X_1 X_{3} = c_1^Y c_2^X c_2^Y c_3^X$, though the worst case/average case is $O(n)$ {\color{edits} in the Majoranas}. Thus, to build a product state requires at best case $O(n)$ number of interacting fermionic gates, which we cannot simulate perturbitavely. \\
\indent These problems may map to circuits which require {\color{edits} greater than logarithmic} scaling number of interacting fermionic gates

\noindent \textit{Proof of statement 3.} 
    We will show that the simulation of $CZ$ conjugated matchgate circuits {\color{edits}implies} the strong simulation of matchgates on magic state inputs. From previous work, we know that strong simulation would imply we could simulate {\color{edits} post selected universal quantum computation, and collapse the polynomial hierarchy to its third level \cite{MatchwOutside, Bremner_2010}}. \\
    To do this, we will show that there exists a way to implement the $|GHZ_4\rangle$ state using product state inputs and $CZ$ conjugated matchgate circuits. It is known that the fermionic state $|GHZ_4\rangle$, and any pure fermionic non-Gaussian state can be used as a magic state for matchgate circuits \cite{NonGMagic}. It is also known that magic states like $|GHZ_4\rangle$ can sometimes be made from two non-fermionic states. We will provide a construction of $|GHZ_4\rangle$ that requires $6$ qubits. \\
    \indent For {\color{edits} the} magic states, let us start with the six qubit product state input $|+\rangle^{\otimes 6}$. We will begin with dividing {\color{edits} the} six qubits into two systems of three, and applying the circuit $G(H^*, H^*)_{23} G(H^*, H^*)_{12} CZ_{12} CZ_{23}$, for 
\begin{equation*}
    H^* = \frac{1}{\sqrt{2}} \begin{pmatrix}
        1 & -1 \\
        1 & 1
    \end{pmatrix}
\end{equation*}
and where we choose {\color{edits} the} $CZ$ circuit to be the one which applies $CZ_{12} CZ_{23}$ for all necessary blocks of 3 input qubits. We then consider {\color{edits} the} full system of $6$ qubits, and apply $G(H,H)_{34}$. After applying fSWAPs to re-arrange qubits 3 and 4 with 5 and 6, and applying the the $CZ^{\dagger}$ circuits to fully conjugate {\color{edits} the} matchgate circuit, we find that measuring the final two qubits to be in state zero results in {\color{edits} the} ancilla register being in $|GHZ_4\rangle$. Thus, all proofs from previous work can be used with the slight modification of not dealing with the $|GHZ_4\rangle$ state directly, but instead measuring two more qubits to be in state $|00\rangle$ for each magic state used. Diagrammatically, this gadgetization that makes magic states can be seen in Fig. \ref{fig:Setup}.  \\

\begin{figure*}
\begin{center}
\begin{adjustbox}{width=0.7\textwidth}
    \begin{quantikz}
        \lstick{\ket{+}} & & \gate[2]{CZ} & \gate[2]{G_{H^*}} & & & & \gate[2]{CZ} & &  \\
        \lstick{\ket{+}} & \gate[2]{CZ} & & & \gate[2]{G_{H^*}} & & & & \gate[2]{CZ} & \\
        \lstick{\ket{+}} & & & & & \gate[2]{G_H} & \gate[4]{fSWAPs} & & & \\
        \lstick{\ket{+}} & & \gate[2]{CZ} & \gate[2]{G_{H^*}} & & & & \gate[2]{CZ} & & \\
        \lstick{\ket{+}} & \gate[2]{CZ} & & & \gate[2]{G_{H^*}} & & & & \gate[2]{CZ} & \rstick{0} \\
        \lstick{\ket{+}} & & & & & & & & & \rstick{0} \\
    \end{quantikz}
\end{adjustbox}
\end{center}
\caption{Circuit for creating $|GHZ_4\rangle$ with $CZ$ conjugated matchgates acting on product state inputs. At the end of the circuit, measuring the final two qubits to be both in the zero state projects the remaining qubits into the $|GHZ_4\rangle$ state. Note that for the utilization of this gadgetization, we can move the matchgates which would utilize this resourceful state to be before we apply the final $CZ$ gates, after the series of fSWAPs. As $CZ$ does not change the state being projected to $|GHZ_4\rangle$ with measuring the final two qubits to be in the zero state, we can still use this is a magic state gadget. }
\label{fig:Setup}
\end{figure*}

\subsection{Clifford Permutation Conjugated Matchgates}

\noindent \textbf{Theorem 3.} \label{thm:CPC} 
    For any Clifford permutation conjugated matchgate circuit, we have at least efficient CIbO (computational input, restricted bitstring output) simulation.

\begin{proof}
\indent For bitstring outputs over all qubits, we can calculate $p(y^*)$ in polynomial time as follows: we consider the calculation of 
\begin{equation}
    p(y^*) = \langle x| C_{\text{P}}^{\dagger} U_{\text{MG}}^{\dagger} C_{\text{P}} | y^* \rangle \langle y^*| C_{\text{P}}^{\dagger} U_{\text{MG}} C_{\text{P}} |x\rangle {\color{edits}.} 
\end{equation}
We know that $C_{\text{P}}|x\rangle = |x ' \rangle$; a Clifford permutation (Clifford with no Hadamard \cite{Bravyi_2021}) maps a bitstring $x$ to some new bitstring $x ' $, and the mapping can be done in polynomial time. We also know that for $|y^*\rangle$, when $y^*$ is a bitstring over all qubits which the Clifford acts on, it is mapped to another bitstring $y^{**}$ on the same bits for the same reason ($C_{\text{P}}|y^*\rangle = |y^{**} \rangle$). This mapping can be done in polynomial time (e.g. due to the simulability of Cliffords on computational basis states). Thus, the problem above can be re-written as 
\begin{equation}
    p(y^*) = \langle x ' | U_{\text{MG}}^{\dagger} | y^{**} \rangle \langle y^{**} | U_{\text{MG}} |x ' \rangle {\color{edits},} 
\end{equation}
and thus we have reduced {\color{edits} the} problem to a matchgate circuit, for which can evaluate this output in polynomial time.  \\
\end{proof}

For attempting to get marginals over outputs $k \not = n$ qubits, attempting to find a reduction back to a matchgate circuit is assumed to be harder. \\
\indent Take for example a marginal over the first $6$ qubits of {\color{edits} the} system being in bitstring $001001$, and let the Clifford permutation we're considering be defined as $C_{\text{P}} = \text{CNOT}_{76}\text{CNOT}_{85}\text{CNOT}_{54}...\text{CNOT}_{21}$. The projector $|y^{**} \rangle\langle y^{**} | \otimes I$ gets mapped to the sum over four distinct projectors, as 
\begin{align*}
    &C_{\text{P}} (|001001\rangle\langle 001001| \otimes I) C_{\text{P}}^{\dagger} \\  
    &= (|11100100\rangle \langle 11100100| + |11100010\rangle \langle 11100010|\\
    &+|00011101\rangle \langle 00011101|+|00011011\rangle \langle 00011011|) \otimes I{\color{edits}.} 
\end{align*}
For some bitstring over the $k$ first qubits in {\color{edits} the} system, and a CNOT circuit of the form (with $k' < k$)
\begin{align*}
     C_{\text{P}} = &(\text{CNOT}_{k+1,k}\text{CNOT}_{k+2,k-1}...\text{CNOT}_{k+k'+1,k-k'}\\
     &\text{CNOT}_{k-k',k-k'-1}\text{CNOT}_{k-k'-1,k-k'-2}...\text{CNOT}_{2,1}){\color{edits},} 
\end{align*}
our single projector is mapped to a sum of up to $2^{k'}$ projectors of the form $|y^{**}  \rangle \langle y^{**}  | \otimes I$. Thus, attempting to map this problem back to a matchgate circuit may require up to an exponential scaling number of Pfaffians to be calculated, and thus be hard to simulate classically. We can also consider the conjugation of a single projector to a sum of projectors as a consequence of the loss of $Z_i \in \mathcal{L}_{2}$, and the lack of preservation of number operator. \\

It is still an open question to find a reduction from this problem to some known hard computational problem. The lack of ability to map back to matchgates however implies that calculating marginals is classically intractable. The exponential number of projectors for which probabilities need to be evaluated resembles problems in circuits which we know are hard to classically sample, such as Boson Sampling tasks \cite{bs1, diag1}. \\
\indent We also point out that in the proof of Theorem \hyperref[thm:CPC]{3}, we can relax the assumption our permutation is a Clifford permutation without the proof changing. We can instead consider conjugation by any generalized permutation, with any phase \cite{jozsa2024iqpcomputationsintermediatemeasurements}. 

\subsection{Clifford Conjugated Matchgate Circuits}

\noindent \textbf{Theorem 4.} \textit{
For an arbitrary Clifford $C$, {\color{edits} reduction to a matchgate problem does not necessarily exist. }} \\

\begin{proof}
    \indent Even on a computational basis state input no simulability conditions should exist, based on the fact that computational basis state inputs are no longer Gaussian states with respect to {\color{edits} the} conjugated circuit. Let us take the Hadamard-layer Clifford (such that we conjugate by a Hadamard on each qubit), $C_{\text{H}}^\dagger = C_{\text{H}} \equiv H_1 \otimes H_2 \otimes ... \otimes H_n$. On the zero state, we have the time evolution task of 
\begin{align*}
    &\langle \mathbf{0}| C_{\text{H}} U_{\text{MG}} C_{\text{H}}|\mathbf{0}\rangle \langle \mathbf{0}|C_{\text{H}} U_{\text{MG}}^{\dagger} C_{\text{H}}|\mathbf{0} \rangle \\
    &= \langle \mathbf{0}| C_{\text{H}} U_{\text{MG}}C_{\text{H}}\left(\prod_{i=1}^{n} \frac{1+Z_i}{2}\right)C_{\text{H}} U_{\text{MG}}^{\dagger}C_{\text{H}} |\mathbf{0} \rangle \\
    &= \langle \mathbf{0}| C_{\text{H}} U_{\text{MG}}\left(\prod_{i=1}^{n} \frac{1+X_i}{2}\right)U_{\text{MG}}^{\dagger} C_{\text{H}} |\mathbf{0} \rangle \\
    &= \langle \mathbf{0}| C_{\text{H}} U_{\text{MG}}\left(\frac{1-c_1}{2}\right)...\left(\frac{1-c_1...c_{2n-1}}{2}\right)U_{\text{MG}}^{\dagger} C_{\text{H}} |\mathbf{0} \rangle\\
    &= \langle \mathbf{0}| \left(\frac{1-c_1^{*'}}{2} \right)...\left(\frac{1-c_1^{*'}c_2^{*'}c_3^{*'}...c_{2n-1}^{*'}}{2}\right) |\mathbf{0} \rangle {\color{edits},} 
\end{align*}
for $c_j^{*'} =  (\sum_{k}R_{jk}C_{\text{H}}c_kC_{\text{H}})$
which is not simulable in general. Thus, {\color{edits} the ability} to simulate bitstring outputs of these states no longer applies.
\end{proof}

{\color{edits}
Proving the computational hardness of this result however, at this time, is not possible. The conjugation prevents mapping to known hard circuits such as IQP, Clifford-magic circuits, or universal circuits \cite{ConjbCliff, Bremner_2010}. However, we conjecture computational hardness, for a number of reasons. One argument goes as follows: let us compose some circuit into $U_2 U_1 |0\rangle$, and assume that there exists a proof that this circuit is hard to calculate marginals from. As a concrete example, we can take $U_1$ to be the Gaussian circuit that makes a product state and $U_2$ to be an arbitrary Clifford. The efficient simulability of probability marginals on Clifford conjugated matchgates would imply that for universal circuits composed of $U_2 U_1 |0 \rangle$, that $U_2 U_1 U_2^{\dagger} | 0 \rangle$ was easy to sample from. Thus, the act of one component of the universal circuit beforehand would have to remove all complexity from the universal problem. This is assumed to be computationally hard.}

\section{Proofs of Complexity of Cliffords following Matchgate Circuits}\label{sec:Expectation}

{\color{edits} We now move from the bitstring output complexity for conjugated circuits, to the Pauli expectation output complexity for Cliffords following matchgates. From our work above as well as previous work on Pauli outputs from Jozsa and Miyake \cite{JozsaMiyake}, we already know about restrictions on Pauli expectation value outputs when the circuit is conjugated.}\\
\indent Rather than relying on the language of matchgates, it will be better for the next Section to discuss {\color{edits} covariance matrices, which fully determine Gaussian states}. \\
\indent To make this connection, we write the evolution of an initial density matrix by a fermionic Gaussian unitary as 
\begin{align*}
    U_{\text{G}} | \mathbf{0}\rangle \langle \mathbf{0}| U_{\text{G}}^{\dagger} &= U_{\text{G}} \left(\prod_{i=1}^n \frac{1-c_{2i-1}c_{2i}}{2}\right) U_{\text{G}}^{\dagger} \\
    &= \prod_{i=1}^n \frac{1-c_{2i-1}'c_{2i}'}{2} {\color{edits},} 
\end{align*}
for $c_i ' = \sum_{k=1}^{2n} R_{i,k}c_k$, $R \in O(2n)$. We have $R \in O(2n)$ rather than $SO(2n)$ to reflect that $O$ can act over the even or the odd parity sector. More importantly, we see that in this framework, Gaussian fermionic states have a compact representation. \\
\indent {\color{edits} Namely, we represent a state by its $2n \times 2n$} covariance matrix
\begin{equation}
    [C(\rho)]_{jk} = -\frac{i}{2}\text{Tr}([c_j, c_k]\rho)
\end{equation}
for $j,k\in[1,2n]$. For a state evolved by a fermionic Gaussian unitary, we have 
\begin{equation}
    C\left(U_{\text{MG}}\rho U_{\text{MG}}^{\dagger}\right) = RC(\rho)R^{\dagger}
\end{equation}
for $R \in O(2n)$. For a fermionic Gaussian state $|\psi\rangle = U_{\text{G}}|\mathbf{0}\rangle$ we can write 
\begin{equation}
    C(|\psi\rangle\langle\psi|) = R \bigoplus_{j=1}^{n} \begin{pmatrix}
        0 & 1 \\ -1 & 0
    \end{pmatrix} R^{T}
\end{equation}
 We can calculate all expectation values over Majoranas, and thus all Pauli strings, with the covariance matrix and Wick's Theorem \cite{PhysRev.80.268}
\begin{equation}\label{Pff}
    (-i)^k \text{Tr}(c_{\mu_1}c_{\mu_2}...c_{\mu_{2k}}\rho) = \text{Pf}(C(|\psi\rangle\langle\psi|)_{\hat \mu \hat \mu}) {\color{edits},} 
\end{equation}
for $C(|\psi\rangle\langle\psi|)_{\hat \mu \hat \mu}$ the submatrix over indices $\hat \mu$. 
For all expectation values of products of an odd number of Majoranas, {\color{edits} the} expectation value {\color{edits} is} zero. Physically we can understand it by recalling that matchgates/free fermion rotations are parity conserving. When taking the expectation value of any Pauli which flips parity, such as $XYX$, {\color{edits} the} expectation value is zero. We will define the set $P_p$ of all parity preserving Paulis, such as $ZXXZ$, and the set $P_b$ of all parity breaking Paulis, such as $XYX$. We define
\begin{align}
    P_p &\equiv \{\, p \ | \ p \in \mathcal{P}_n, \ 
    \text{parity}(p|m\rangle) = \text{parity}(|m\rangle) \}  {\color{edits},} \\[0.5em]
    P_b &\equiv \{\, p \ | \ p \in \mathcal{P}_n, \ 
    \text{parity}(p|m\rangle) = (\text{parity}(|m\rangle)+1)\}{\color{edits},}
\end{align}
with binary addition over $\mathbb{F}_2$, the field over integers $\{0,1 \}$, {\color{edits}and} the parity of some bitstring $|m\rangle$ as $\text{parity}(|m\rangle)$. \\
\indent The operators $c_{\mu_{\text{I}}}$ in equation \eqref{Pff} are the Jordan-Wigner Majoranas, and the covariance matrix stores the expectation values of all quadratic products of {\color{edits} the} Jordan-Wigner Majoranas. As a concrete example, submatrix over indices $\hat \mu = \{1,2\}$, $\text{Pf}(C(|\psi\rangle\langle\psi|)_{\hat \mu \hat \mu})$ tells us about the $Z_1$ expectation value, as the entry corresponding to the expectation value of $-i c_1 c_2 = -i X Y = Z$. \\ 
\indent With this understanding, we can extend the use of {\color{edits} the} covariance matrix to consider both Cliffords acting on matchgate circuits, and Cliffords acting on matchgate circuits with access to dynamics from linear fermionic terms. 

\subsection{Simulability of Cliffords following Matchgate Circuits on Computational Basis State Inputs}\label{sec:CIPO}

\noindent \textbf{Theorem 5.} \label{thm:CIPO} \textit{
    For any circuit of the form $C^{\dagger}U_{\text{MG}}$, we have at least efficient CIPO simulation {\color{edits} via $2n \times 2n$ covariance matrices}}.
\\

\begin{proof}
    Rather than asking about the covariance matrix with respect to the Jordan-Wigner Majoranas, we can ask about the covariance matrix with respect to the encoding given by $C^{\dagger} \{ \mathcal{L}_{1}^{\text{(JW)}} \} C$. Let the Majoranas of this encoding be $C^{\dagger}c_i C = c_i'$. Studying the covariance matrix, for our Clifford conjugated encoding, we find 
\begin{align*}
    [C(\rho)]_{jk} &= -\frac{i}{2}\text{Tr}([c_j', c_k']C^{\dagger} U_{\text{MG}} |\mathbf{0}\rangle\langle\mathbf{0}| U_{\text{MG}}^{\dagger} C) \\
    &= -\frac{i}{2}\text{Tr}(C^{\dagger} [c_j, c_k]C C^{\dagger}U_{\text{MG}} |\mathbf{0}\rangle\langle\mathbf{0}| U_{\text{MG}}^{\dagger} C) \\
    &= -\frac{i}{2}\text{Tr}([c_j, c_k]U_{\text{MG}} |\mathbf{0}\rangle\langle\mathbf{0}| U_{\text{MG}}^{\dagger}) {\color{edits},}
\end{align*}
which is exactly the covariance matrix of our circuit without Clifford conjugation. Thus, for Clifford conjugated circuits, we can understand {\color{edits} the} Gaussian fermionic states as being represented by covariance matrices defined over a different set of $2n$ anticommuting operators. For every expectation value of {\color{edits} the} conjugated circuit, there exist a corresponding expectation value in {\color{edits} the} unconjugated circuit. Thus, for any circuit of the form $C^{\dagger}U_{\text{MG}}$, we have at least efficient CIPO simulation. \\
\end{proof}

\indent When we have Cliffords following matchgates acting on a computational basis state input, we still have that half of our Pauli expectation values are zero. We can consider an encoded set of parity preserving and parity breaking operators, $C P_p C^{\dagger} = P_p'$ and $C P_b C^{\dagger} = P_b'$,  and have that $\langle p \rangle = \langle \mathbf{0}|U_{\text{MG}}^{\dagger} C^{\dagger}pC^{\dagger}U_{\text{MG}}|\mathbf{0}\rangle = 0$ for $p \in P_b'$. 

\subsection{Simulability of Cliffords following Matchgate Circuits on Product State Inputs}

\noindent \textbf{Theorem 6.}\label{thm:PIPO} \textit{
    For any circuit of the form $C^{\dagger}U_{\mathcal{L}_{1,2}}|\mathbf{0}\rangle$, where $U_{\mathcal{L}_{1,2}}$ is generated from a Hamiltonian with quadratic and linear products of terms from {\color{edits} the} Jordan-Wigner Majoranas, we have at least efficient PIPO simulation {\color{edits} via a $(2n+1) \times (2n+1)$ covariance matrix.}}

\begin{proof}
    We will first consider having no Clifford conjugation, and just discuss adapting the covariance matrix framework to include dynamics from linear Majorana terms. The key insight is that we can extend {\color{edits} the} set of operators by {\color{edits} the} parity operator, as we did in Lemma \hyperref[cor:ExtL]{1}, and consider the evolution of a covariance matrix of size $(2n+1) \times (2n+1)$. We can define our new covariance matrix  
\begin{equation*}
    [C(\rho)]_{jk} = -\frac{i}{2}\text{Tr}([d_j, d_k]\rho)
\end{equation*}
for {\color{edits} $d_i$ defined as the extended encoding of Lemma \hyperref[cor:ExtL]{1}, and $j,k\in[2n+1]$}. When $j,k \not=0$ we have that $d_j d_k = c_j c_k$, and thus it is only when one index is zero that we are examining the expectation values of {\color{edits} the} single Majorana terms. For $O = e^{4H_d} \in O(2n+1)$, with $H_d$ {\color{edits} the} Gaussian Hamiltonian over {\color{edits} the} extended encoding, we have 
\begin{equation*}
    C(U_{\mathcal{L}_{1,2}}\rho_0 U_{\mathcal{L}_{1,2}}^{\dagger}) = OC(\rho_0)O^{\dagger} {\color{edits},}
\end{equation*}
where 
\begin{equation}
    C(\rho_0) = \begin{pmatrix}
        0 & 0 & 0 & 0 & 0 & ... \\
        0 & 0 & 1 & 0 & 0 & ... \\
        0 & -1 & 0 & 0 & 0 & ... \\
        0 & 0 & 0 & 0 & 1 & ... \\
        0 & 0 & 0 & -1 & 0 & ... \\
        \vdots & \vdots & \vdots&  \vdots & \vdots & \ddots 
    \end{pmatrix}{\color{edits}.}
\end{equation}

For getting Pauli expectation values, we notice that with the addition of the parity operator to {\color{edits} the} system (which for Jordan-Wigner is $ZZZ...ZZ$), we can write any Pauli as two possible products of Majoranas rather than a unique product of Majoranas. For Pauli strings in $P_p$, there exist an even product of Majoranas which doesn't include the parity operator, and an odd product which does. For $P_b$, vice versa. With the ability to write \textit{any} Pauli as a product of an even number of Majoranas, we can calculate nonzero expectation values even over Pauli strings which break parity. For any expectation value over an operator in $P_p$, we can write the expectation value as the Pfaffian over the reduced matrix which does not include row/column zero. For any expectation value in $P_b$ however, we can write the expectation value as an even product of Majoranas, which must include row/column zero. Specifically, 
\begin{equation}
\begin{aligned}
     -i^k\text{Tr}(d_0 d_{u_1}...d_{u_{2k-1}} \rho) &= -i^k\text{Tr}(c_0 i c_{u_1} c_0 c_{u_2}... c_{u_{2k-1}}\rho) \\
     &= -i^{k+1}\text{Tr}( c_{u_1} c_{u_2}... c_{u_{2k-1}}\rho){\color{edits}.}
\end{aligned}
\end{equation}\\
We can simply extend the proof from Section \ref{sec:CIPO}, and find that for $C^{\dagger}U_{\mathcal{L}_{1,2}}|\mathbf{0}\rangle$, we can calculate all Pauli expectation values in polynomial time. \\ 
\end{proof}

\indent It is clear that the simulability of these circuits contains the simulability of Cliffords on product states. With access to $U_{\mathcal{L}_{1,2}}$ we can make product state inputs without the Hadamard device. Thus, we can consider these circuits to be generalizations of the simulability of Cliffords on product state, extending the class of input states such that we can calculate all Pauli expectation values to be all states of the form $U_{\mathcal{L}_{1,2}}|\mathbf{0}\rangle$. \\
\indent Code accompanying Theorems \hyperref[thm:CIPO]{5} and \hyperref[thm:PIPO]{6} is available at \url{https://github.com/andrewprojansky/Clifford-Matchgate-Hybrids}. 
\\
\\
\textit{A note on covariance matrices of size $(2n+2) \times (2n+2)$}: We discuss extending the covariance matrix by adding the total parity operator to {\color{edits} the} set of Majoranas. On face value this may seem like it has two possible problems. 
\begin{enumerate}
    \item Pauli expectation values are now represented with two possible products of Majoranas.
    \item The reduction of expectation value to the Pfaffian relies on Wick's Theorem \cite{PhysRev.80.268}. For our parity operator, as it comes from a single Majorana operator rather than a linear combination of raising/lowering operator, it is not obvious that we can still use Wick's Theorem.
\end{enumerate}
There is a way to sidestep both of these problems however, if instead of using the total parity operator to extend our matchgates, we follow the procedure in Knill in Lemma 2 \cite{knill2001fermionic}. Instead of a single Majorana, we add two Majoranas, by creating an isomorphism from $\mathcal{L}_{1,2}^{(n)}$ in {\color{edits} the} original space over $n$ qubits to $\mathcal{L}_{2}^{(n+1)}$ in a space of $n+1$ qubits. \\
\indent We thus have a physical space of Pauli strings, and a logical space which it is encoded into. Our logical space consist of all Pauli strings over $n$ qubits. For each Pauli $p$ in {\color{edits} the} logical space, 
\begin{itemize}
    \item If the $p$ is parity preserving, then $p$ in the logical space is $I \otimes p$ in the physical space. \item If $p$ is not parity preserving, the $p$ in the logical space is $X \otimes p$ in the physical space. 
\end{itemize}
Any parity preserving operator is mapped to itself, and any parity breaking operator is mapped to a parity preserving operator over an additional qubit. \\
\indent Whenever we want to evolve by a linear fermionic operator $c_i$ in {\color{edits} the} logical space, we evolve by quadratic operator $X \otimes p$ in {\color{edits} the} physical space. If we want to take the expectation value over an odd product of Majoranas $p$ in {\color{edits} the} encoded space, we instead take the expectation value over $X \otimes p$. \\
\indent For {\color{edits} the} new extended encoding Jordan-Wigner Majoranas $ \{ c_1', c_2', ... \} \in \mathcal{L}_{1}^{(n+1)}$, $X \otimes p_i = c_2' ...$ always includes $c_2'$ when written as a product of Majoranas, while never including $c_1'$. Thus, any rotation matrix in $SO(2n+2)$ has trivial first row/column. For any covariance matrix in our physical space, we can write
\begin{equation}
    C(\rho) = \begin{pmatrix}
        0 & 1 & 0 & 0 & ... \\
        -1 &  &   &   & \\
        0 &  &   &\mathbf{C'(\rho)}   & \\
        \vdots &  &   &   & \\
    \end{pmatrix}{\color{edits},}
\end{equation}

for $C'(\rho)$ the covariance matrix over $2n+1$ Majoranas.  \\

\subsection{Expectation Values and Physical Intuition of Clifford Conjugated Matchgate Circuits}

\noindent \textbf{Jozsa and Miyake [2008]} \textit{
        For arbitrary Clifford conjugated matchgate circuits, we have at least efficient PIpO simulation.}
\\
\begin{proof}
    For any encoding, there exist $O(\textit{poly}(n))$ amount of spins which can be written as $p = \prod_{i=1}^d c_{j_i}$ for $c_{j_i}$ a Majorana of our conjugated encoding encoding and $d \in O(1)$. For all those operators $p$,
\begin{align*}
    \langle p\rangle &= \langle \psi_{\text{prod}} | C^{\dagger} U_{\text{MG}}^{\dagger} C p C^{\dagger} U_{\text{MG}} C | \psi_{\text{prod}} \rangle{\color{edits},} \\
    &= \langle \psi_{\text{prod}} | C^{\dagger} U_{\text{MG}}^{\dagger} C c_{i_1} c_{i_2}... c_{i_d} C^{\dagger} U_{\text{MG}} C | \psi_{\text{prod}} \rangle{\color{edits},} \\
    &= \sum_{j_1, j_2, ..., j_d} R_{i_1 j_1} R_{i_2 j_2} ... R_{i_d j_d} \langle \psi_{\text{prod}} | c_{j_1} c_{j_2}... c_{j_d} | \psi_{\text{prod}} \rangle {\color{edits},}
\end{align*}
 can be calculated in polynomial time. \\
\end{proof}

\indent Even though these circuits map to calculations which look like they have restricted interacting fermionic gates followed by a sequence of non-interacting fermionic gates, we can still simulate some number of Pauli expectation values in polynomial time. Thus, a lack of stronger simulability does not mean that there is no Pauli expectation value in {\color{edits} the} system which we can't simulate in polynomial time. \\
\indent For example, consider choosing a pruned Sierpinski encoding \cite{harrison2024sierpinskitriangledatastructure,HarrisonSerp} over ten qubits, in which we have fermionic operators with the same locality as the Ternary Tree encoding \cite{Jiang_2020}. The encoding defines the set of operators
\begin{align*}
    C_1 = XXIIXIIIII{\color{edits},} \; \; & \; \; C_2 = ZXIIXIIIII{\color{edits},} \\
    C_3 = IYYIXIIIII{\color{edits},} \; \; & \; \; C_4 = IZIXXIIIII{\color{edits},} \\
    C_5 = IZIZXIIIII{\color{edits},} \; \; & \; \; C_6 = IIIIYYIZII{\color{edits},} \\
    C_7 = IIIIYIXYII{\color{edits},} \; \; & \; \; C_8 = IIIIYIZYII{\color{edits},} \\
    C_9 = IIIIYIIXYI{\color{edits},} \; \; & \; \; C_{10} = IIIIZIIIIX{\color{edits},} \\
    C_{11} = YXIIXIIIII{\color{edits},} \; \; & \; \; C_{12} = IYZIXIIIII{\color{edits},} \\
    C_{13} = IYXIXIIIII{\color{edits},} \; \; & \; \; C_{14} = IZIYXIIIII{\color{edits},} \\
    C_{15} = IIIIYZIZII{\color{edits},} \; \; & \; \; C_{16} = IIIIYXIZII{\color{edits},} \\
    C_{17} = IIIIYIYYII{\color{edits},} \; \; & \; \; C_{18} = IIIIYIIXZI{\color{edits},} \\
    C_{19} = IIIIYIIXXI{\color{edits},} \; \; & \; \; C_{20} = IIIIZIIIIY{\color{edits}.} \\
\end{align*} \\
\indent For this encoding, local interactions on qubit $1$ require quadratic products of Majoranas. However, the operations needed to build fSWAPs requires products of a product of four Majoranas for best case, and at worst 10 Majoranas with the fermionic interactions required to perform fSWAP from qubits $1$ to $5$, and $5$ to $10$. We leave the details of building these operators as an exercise to the reader. \\
\indent Going through this example, we find that the expectation values of the form 
\begin{equation*}
    \langle p\rangle = \langle \psi_{\text{prod}} | C^{\dagger} U_{\text{MG}}^{\dagger} C p C^{\dagger} U_{\text{MG}} C | \psi_{\text{prod}} \rangle 
\end{equation*}

can be thought of {\color{edits}as} a circuit acting on a Gaussian computational basis state $|\mathbf{0}\rangle$ which first has a select number of interacting fermionic gates followed by a free fermion circuit
\begin{equation}
     \langle \mathbf{0} | U_{\text{I}}^{\dagger} U_{\text{F}}^{\dagger} p U_{\text{F}} U_{\text{I}} | \mathbf{0} \rangle{\color{edits},}
\end{equation}
where $U_{\text{I}}|\mathbf{0}\rangle = |\psi_{\text{prod}}\rangle$

\section{Clifford and Matchgate Simulability}\label{sec:Sim}

{\color{edits} In some sense, the understanding that Cliffords following matchgates retain simulable expectation values is simple; the action of the Clifford can be absorbed into the Paulis, and for any matchgate circuit all Pauli expectation values are efficiently simulable. Thus, its important to ask what this explicit construction accomplishes.
It is worth taking some time to highlight the consequences of our work on Cliffords following matchgates and our understanding of simulability.\\
{\color{edits} We define \textit{Gaussianity} based on conditions derived from Brayvi; a state $|\psi\rangle$ is Gaussian if and only if \cite{bravyi2004lagrangianrepresentationfermioniclinear}

\begin{equation} \label{eq:Gauss}
    \sum_{i=1}^{2n} c_k \otimes c_k |\psi\rangle |\psi\rangle = 0 {\color{edits}.}
\end{equation}

No specification is made in regards to some encoding being used in this definition, though in practice this definition is used explicitly with the Jordan-Wigner encoding. We can thus consider extending the definition of \textit{Gaussianity} to different fermionic encodings. \\
We quickly see that we cannot extend Bravyi's definition to completely arbitrary fermionic encodings. For any state $|\psi\rangle = U|00..0\rangle$, there exist an encoding $U c_{i_{JW}}U^{\dagger} = c_i^*$ such that

\begin{align*}
	\sum_{k=1}^{2n} c_k^* \otimes c_k^* |\psi\rangle |\psi\rangle &= (U \otimes U) \sum_{k=1}^{2n} c_{k_{JW}} \otimes c_{k_{JW}} (U^{\dagger} \otimes U^{\dagger}) (U \otimes U)  |00...0\rangle |00...0\rangle \\
&= (U \otimes U) \left[ \sum_{i=1}^{2n} c_{k_{JW}} \otimes c_{k_{JW}} |00...0\rangle |00...0\rangle \right] \\
&= 0
\end{align*}

If we allow for arbitrary complexity within our Majoranas, we can argue \textit{any} state is Gaussian.
}

Thus, we must make some restriction to what we allow for our Majoranas such that {\color{edits} the} Gaussian circuits remain simulable. We take the perspective that so long as our Majoranas are single Pauli strings, {\color{edits} the} system is classically simulable. Thus, explicitly recognizing the simulability of Cliffords following matchgates is a recognition of \textbf{all states in Hilbert space that are Gaussian with respect to reasonable choice of fermionic operators}. These states necessarily go beyond stabilizers or matchgate states; they have varied operator structure, entanglement entropy, and are not even guaranteed to be close to one another in distance metrics on quantum states. Thus, we have extended the set of simulable quantum states. We hope that in future approximate methods are benchmarked against \textit{any} of these simulable Gaussian states, to truly highlight when a problem has left the Gaussian manifold.
}

\indent This perspective {\color{edits} is also useful for a fundamental reason}; it lets us see all stabilizer states as Gaussian fermionic states. {\color{edits} 
If we allow our Majoranas to not be Jordan-Wigner but some other set of mutually anticommuting Paulis, we can always find a choice such that for any stabilizer state this condition holds.} \\ \\
{\color{edits}
\noindent \textbf{Corollary 1.} \textit{For any stabilizer states $|\psi_{stab}\rangle$ there exist some mutually anticommuting set of Paulis $c_i \in \mathcal{L}_{1}^{n}$ such that Equation \ref{eq:Gauss} is true. } }
\\

\indent We can write {\color{edits} any stabilizer} state with a vacuum covariance operator, where $c_{2i-1}'c_{2i}'$ are the stabilizers of our state. We can understand all stabilizer states by considering all fermionic vacuum covariance matrices, with respect to different sets of mutually anticommuting Pauli strings. {\color{edits}We can thus identity the simulability of stabilizer states via their Gaussian structure; all Pauli expectation values can be computed from Pfaffians of covariance matrices via Wicks theorem, with the recognition that stabilizers are generated by quadratic products of Majoranas. }

\section{Conclusion}\label{sec:Conclusion}
In this paper, we have extended the simulability of both matchgate and Clifford circuits by considering Clifford/matchgate hybrid circuits. We have shown that for {\color{edits} conjugated matchgates, the complexity of getting bitstring outputs can be understood via how we spread fermionic information through qubit space.} When considering conjugated matchgates, we are considering the time dynamics of free fermion solvable spin systems \cite{chapman2023unified,Chapman_2020, Elman_2021}, {\color{edits} and thus also show when dynamics free fermion solvable spin Hamiltonians are easy to sample from or not}.  For Pauli expectation values, we have shown that we can extend the family of input states such that Clifford evolution retains all Pauli expectation values to be any matchgate input state, rather than just product states. {\color{edits} In doing so, we clarify the full space of states which are \enquote{Gaussian}.}\\
\indent We now turn towards discussing future directions. It would be interesting to study possible uses for these circuits in classical shadow tomography \cite{Huang_2020, Wan_2023}. As we present circuits such that all Pauli expectation values are classically simulable, we wonder about their utility as the family of simulable circuits chosen in the protocol. 
%it would be interesting to see whether by considering subsets of Cliffords and subsets of matchgates, if we can outpeform the cost of the algorithm for those subsets individually. \\

\indent Another question we could ask is about how these new circuits are distributed throughout Hilbert space. It is well known that Clifford circuits form a $3$-design, while matchgates do not even form a $2$-design \cite{Wan_2023, Liu2018, webb2016cliffordgroupformsunitary}. It is also known that combining different simulable circuits together can enhance the complexity of quantum states \cite{PhysRevLett.123.170502, lami2024quantumstatedesignsclifford}. It would be interesting to study the complexity of some of the circuits introduced here from the perspective of $t$-designs, especially for Cliffords following matchgate circuits. \\
\indent Related, we can also ask about the utility of these circuits in benchmarking applications \cite{burkat2024lightweightprotocolmatchgatefidelity}. We can ask how close circuits we might want to run on a quantum computer are to these new simulable circuits, as an attempt to prove circuits are robust against classical spoofing experiments. \\
\indent The circuits introduced here may also have applicability in optimization problems related to ground state solutions for physical systems of interest. It is known that both Cliffords and Gaussian fermionic states have applicability in classical optimization problems for approximating quantum systems \cite{Schleich_2023, gulania2023hybridalgorithmtimedependenthartreefock, Bravyi_impurity, Bravyi_approximation, Herasymenko_2023}. As our hybrid circuits can be seen as being mixes of Gaussian and non-Gaussian evolution,  we may have access to states which go beyond what we could reach with Hartree-Fock/mean field solvers \cite{Bravyi_impurity, Bravyi_approximation, Herasymenko_2023, henderson2024hartreeFockbogoliubov}. It may also be possible that these circuits motivate better initial states for warm starting quantum variational algorithms, as we can perform some classical variation before letting our quantum computers continue to optimize \cite{Egger_2021, tate2022bridgingclassicalquantumsdp}. \\
\indent Finally, the mapping of Clifford conjugated matchgate circuits to mixed noninteracting and interacting fermionic time dynamics helps explain previous work by the authors in \cite{Projansky_2024} in entanglement spectra statistics. We hope this motivates works on the analytics and theory behind other chaotic signatures such as OTOCs and operator entanglement and their connection to simulability \cite{Hashimoto_2017, Bertini_2020}. \\

\section*{Acknowledgements}
The authors thank Joseph Gibson and Sergii Strelchuk for discussion and suggestions on structure. {AMP, JN, and JDW are supported by the Office of Science, Office of Advanced Scientific Computing Research under programs Fundamental Algorithmic Research for Quantum Computing and Optimization, Verification, and Engineered Reliability of Quantum Computers project. JN and JDW were also supported by the U.S. Department of Energy, Office of Basic Energy Sciences, under Award DE-SC0019374. JDW holds concurrent appointments at Dartmouth College and as an Amazon Visiting Academic. This paper describes work performed at Dartmouth College and is not associated with Amazon.} \\
\section*{Bibliography}

\bibliographystyle{iopart-num}
\bibliography{Pbib2}

\end{document}